\documentclass[12pt,preprint]{aastex}
\usepackage{natbib}
\usepackage{lscape}%,graphix}

\newcommand{\lsim}{~\rlap{$<$}{\lower 1.0ex\hbox{$\sim$}}}
\newcommand{\gsim}{~\rlap{$>$}{\lower 1.0ex\hbox{$\sim$}}}

\slugcomment{To appear in ApJ Letters}

\shorttitle{Imaging the Sun with the MWA Prototype}
\shortauthors{Oberoi et al.}

\begin{document}

\title{
First Spectroscopic Imaging Observations of the Sun at Low Radio Frequencies
with the Murchison Widefield Array Prototype}

\author{Divya Oberoi\altaffilmark{1},
Lynn D. Matthews\altaffilmark{1},
Iver H. Cairns\altaffilmark{2},
David Emrich\altaffilmark{3},
Vasili Lobzin\altaffilmark{2},
Colin J. Lonsdale\altaffilmark{1}, 
Edward H. Morgan\altaffilmark{4},
T. Prabu\altaffilmark{5},
Harish Vedantham\altaffilmark{5},
Randall B. Wayth\altaffilmark{3},
Andrew Williams\altaffilmark{6}, 
Christopher Williams\altaffilmark{4},
Stephen M. White\altaffilmark{7} 
%Builders List\altaffilmark{7}
G. Allen\altaffilmark{8},
Wayne Arcus\altaffilmark{3},
David Barnes\altaffilmark{9},
Leonid Benkevitch\altaffilmark{1},
Gianni Bernardi\altaffilmark{10},
Judd D. Bowman\altaffilmark{11},
Frank H. Briggs\altaffilmark{12},
John D. Bunton\altaffilmark{8},
Steve Burns\altaffilmark{13},
Roger C. Cappallo\altaffilmark{1},
M. A. Clark\altaffilmark{14},
Brian E. Corey\altaffilmark{1},
M. Dawson\altaffilmark{12},
David DeBoer\altaffilmark{8,15},
A. De Gans\altaffilmark{12},
Ludi deSouza\altaffilmark{8},
Mark Derome\altaffilmark{1},
%%Avinash Deshpande\altaffilmark{5}, (opted out)
%Sheperd S. Doeleman,\altaffilmark{1}, (removed from builders list)
R. G. Edgar\altaffilmark{14,16},
T. Elton\altaffilmark{8},
%%Bryan Gaensler\altaffilmark{2}, (opted out)
Robert Goeke\altaffilmark{4},
M. R. Gopalakrishna\altaffilmark{5},
Lincoln J. Greenhill\altaffilmark{10},
Bryna Hazelton\altaffilmark{17},
David Herne\altaffilmark{3},
Jacqueline N. Hewitt\altaffilmark{4},
P. A. Kamini\altaffilmark{5},
David L. Kaplan\altaffilmark{18},
Justin C. Kasper\altaffilmark{10},
Rachel Kennedy\altaffilmark{1,15},
Barton B. Kincaid\altaffilmark{1},
Jonathan Kocz\altaffilmark{12},
R. Koeing\altaffilmark{8},
Errol Kowald\altaffilmark{12},
%%Eric Kratzenberg\altaffilmark{1}, (opted out)
%Deepak Kumar\altaffilmark{5}, (removed from builders list)
Mervyn J. Lynch\altaffilmark{3},
S. Madhavi\altaffilmark{5},
%Michael Matejek\altaffilmark{4}, (removed from builders list)
Stephen R. McWhirter\altaffilmark{1},
Daniel A. Mitchell\altaffilmark{10},
Miguel F. Morales\altaffilmark{17},
A. Ng\altaffilmark{8},
Stephen M. Ord\altaffilmark{10},
Joseph Pathikulangara\altaffilmark{8},
Alan E.~E. Rogers\altaffilmark{1},
Anish Roshi,\altaffilmark{5,19},
Joseph E. Salah\altaffilmark{1},
Robert J. Sault\altaffilmark{20},
Antony Schinckel\altaffilmark{8},
N. Udaya Shankar\altaffilmark{5},
K. S. Srivani\altaffilmark{5},
Jamie Stevens\altaffilmark{8},
Ravi Subrahmanyan\altaffilmark{5},
D. Thakkar\altaffilmark{2},
Steven J. Tingay\altaffilmark{3},
J. Tuthill\altaffilmark{8},
Annino Vaccarella\altaffilmark{12},
Mark Waterson\altaffilmark{3,12},
Rachel L. Webster\altaffilmark{20} and
Alan R. Whitney\altaffilmark{1}
}

\altaffiltext{1}{MIT Haystack Observatory, Westford, MA USA}
\altaffiltext{2}{University of Sydney, Sydney, Australia}
\altaffiltext{3}{Curtin University, Perth, Australia}
\altaffiltext{4}{MIT Kavli Institute for Astrophysics and Space Research, Cambridge, MA USA}
\altaffiltext{5}{Raman Research Institute, Bangalore, India}
\altaffiltext{6}{The University of Western Australia, Perth, Australia}
\altaffiltext{7}{Air Force Research Laboratory, Kirtland, NM USA}
\altaffiltext{8}{CSIRO Astronomy and Space Science, Australia}
\altaffiltext{9}{Swinburne University of Technology, Melbourne, Australia}
\altaffiltext{10}{Harvard-Smithsonian Center for Astrophysics, Cambridge, MA USA}
\altaffiltext{11}{School of Earth and Space Exploration, Arizona State University, Tempe, AZ USA}
\altaffiltext{12}{The Australian National University, Canberra, Australia}
\altaffiltext{13}{Burns Industries, Inc. Nashua, NH USA}
\altaffiltext{14}{Harvard University, Cambridge, MA USA}
\altaffiltext{15}{University of California, Berkeley, CA USA}
\altaffiltext{16}{Massachusetts General Hospital, Boston, MA USA}
\altaffiltext{17}{University of Washington, Seattle, WA USA}
\altaffiltext{18}{University of Wisconsin - Milwaukee, Milwaukee, WI USA}
%\altaffiltext{18}{REU student at MIT Haystack Observatory, University of California, Berkeley, CA USA}
\altaffiltext{19}{National Radio Astronomy Observatory, Green Bank, WV USA}
\altaffiltext{20}{The University of Melbourne, Melbourne, Australia}

%\altaffiltext{9}{Commonwealth Scientific and Industrial Research Organization, Epping, NSW, Australia}
%\altaffiltext{15}{Commonwealth Scientific and Industrial Research Organization, Australia Telescope National Facility, Narrabri, NSW, Australia}
%\altaffiltext{X}{CASS}

\begin{abstract}
We present the first spectroscopic images of solar radio transients
from the prototype for the Murchison Widefield Array (MWA), 
observed on 2010 March~27.
Our observations span the instantaneous frequency band 170.9--201.6~MHz. 
Though our observing period is characterized as a period
of  `low' to `medium' activity, 
one broadband emission feature and numerous short-lived, narrowband, 
non-thermal emission features are evident.
Our data represent a significant advance in low radio frequency solar imaging,
enabling us to follow the spatial, spectral, and temporal
evolution of events simultaneously and in unprecedented detail.
The rich variety of features seen here reaffirms the coronal diagnostic 
capability of low radio frequency emission and provides an early glimpse of 
the nature of radio observations that will become available as the next 
generation of low frequency radio interferometers 
come on-line over the next few years.
\end{abstract}


\section{Introduction}
\label{Sec:intro}
Low radio frequency ($\nu\lesssim300$~MHz) emission provides powerful 
diagnostics of the solar corona. 
However, high fidelity solar imaging at low frequencies is challenging. 
Coronal emission features are complex and dynamic, evolving 
rapidly in space, time, and frequency. 
Consequently, the limited instantaneous spatial and frequency coverage 
provided by current radio interferometers have been inadequate to 
simultaneously resolve transient solar phenomena spatially, temporally, 
and spectrally.

This situation should change dramatically in the next few
years as a new generation of low-frequency radio arrays becomes 
available, leveraging recent advances in digital signal processing 
hardware and computational capacity.
The Murchison Widefield Array (MWA) \citep{Lonsdale09} will be one such
array, with most of its 512 elements spread over 1.5~km 
and a few outliers out to 3~km. 
The resulting dense instantaneous monochromatic $uv$ coverage will provide a radio imaging capability with unprecedented fidelity and flexibility.
%The MWA will have instantaneous monochromatic $uv$ coverage approaching Nyquist sampling providing a radio imaging capability with unprecedented fidelity and flexibility.
The MWA is currently under construction at the Murchison
Radioastronomy Observatory, in the remote and radio-quiet 
Western Australian
outback, and a prototype system comprising 32 interferometer elements 
(tiles) is operational on site. 
This system, hereafter referred to as the ``32T'', 
serves as an engineering testbed and provides early science
opportunities, in advance of the full MWA.
We present here solar  imaging observations obtained with the 32T.  
The quiescent solar emission in the MWA frequency range of 80--300~MHz
is dominated by coronal emission from heights of
$\sim$1--10~R$_{\odot}$ above the photosphere.
These are among the first high-fidelity, high dynamic-range,
spectroscopic images of the Sun with a good temporal and spectral resolution
at meter wavelengths.

\section{Data}
\label{Sec:data}
\begin{figure}
\includegraphics[width=8cm,angle=-90,trim = 0mm 0mm 0mm 0mm, clip=true]{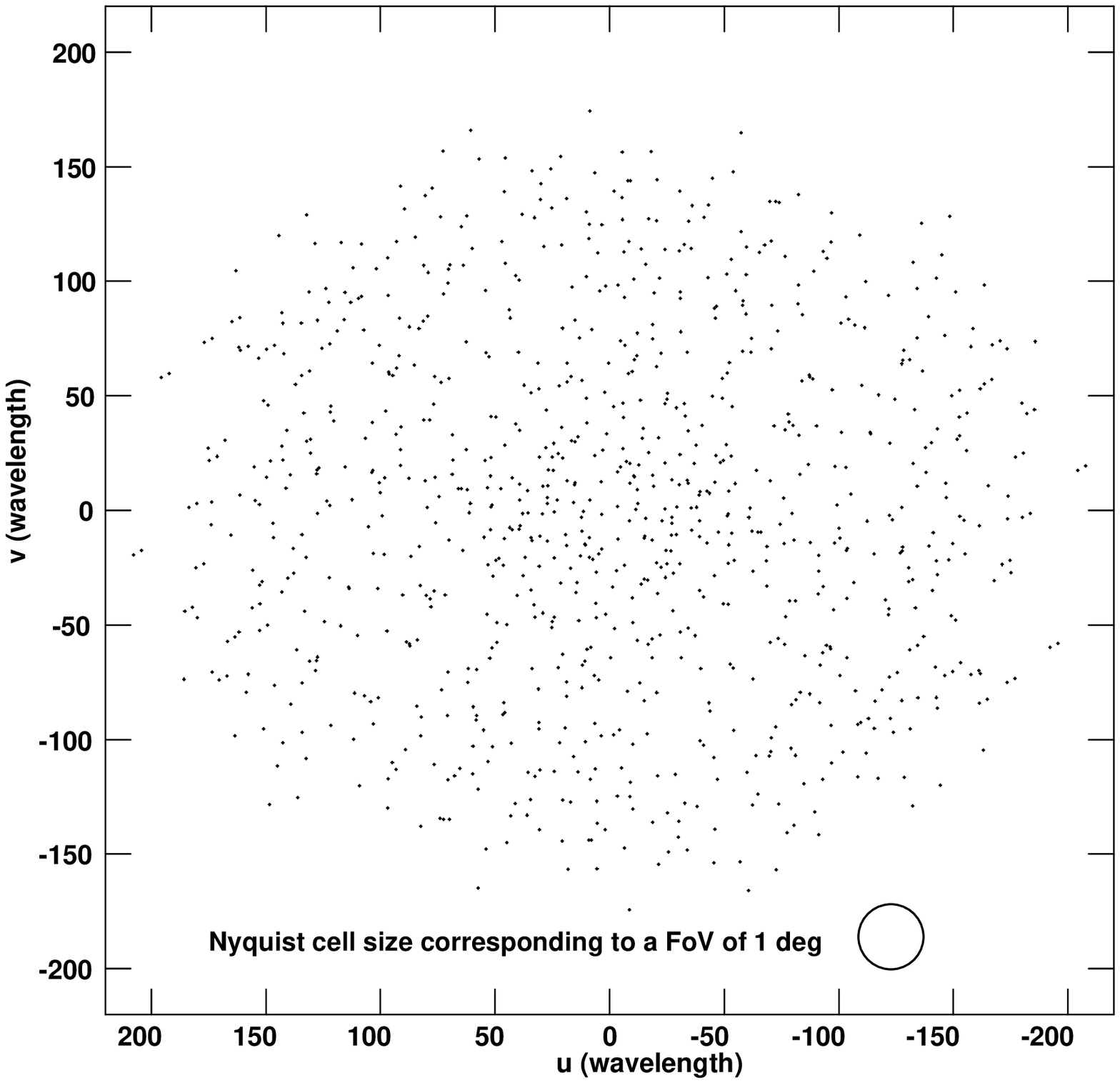}
\caption{
The instantaneous {\em uv} coverage for a single coarse channel 
centered at 186.2~MHz. 
The circle indicates the size of a {\em uv}-cell, 
corresponding to the Nyquist sampling criterion for a 1 degree
FoV.
}
\label{Fig:uvplot}
\end{figure}

The 32T tiles are arranged along the arms of a randomized Reuleaux
triangle (see Cohanim et al. 2004), 
providing a fairly uniformly sampled $uv$ plane with 
baseline lengths up to 350~m (Fig.~\ref{Fig:uvplot}).
%%%%%%%%%%%%%%%%%%%%%%%%%%%%%%%%%%%%%%%%%%%%%%%%%%%%%%%%%%%%%%%%%%%%%%
%%%%%%%%%%%%%%%%%%%%%%%%%%%%%%%%%%%%%%%%%%%%%%%%%%%%%%%%%%%%%%%%%%%%%%%%%%%%
For solar observations, the correlated flux is dominated by the Sun, reducing the effective field-of-view (FoV) to $\sim$1$^{\circ}$, for which the {\em uv} sampling provided by the 32T exceeds the Nyquist criterion\footnote
{$\delta u \sim 1/2\theta_0$, where $\delta u$ is length scale in the {\em uv} plane beyond which visibilities are no longer correlated, and $\theta_0$ is the size of the FoV in radians}. 
This, together with a high signal-to-noise ratio, permits robust, high fidelity imaging \citep{Bracewell_Roberts54}.

In its usual interferometric mode of operation, the 32T 
provides auto- and cross-correlations (visibilities) for all 64 input signals 
(32 tiles$\times$2 linear polarizations).
The entire RF band is directly sampled and filtered into 1.28~MHz wide 
{\em coarse} channels.
Twenty four of these coarse channels are further processed by the correlator,
which can currently provide a time resolution of 50~ms with a 50\% duty 
cycle over a 30.72~MHz band and a spectral resolution of 40~kHz.
The 32T design closely follows the MWA architecture described in 
\citet{Lonsdale09}.

The data presented here were obtained on 2010 March~27 from 04:24:53 
to 04:34:03~UT and span the range 170.8--201.6~MHz. They
were smoothed to a time resolution of 1~second before further processing.
Data editing, calibration, and imaging were performed using the
Astronomical Image Processing System.
Self-calibration, to solve for frequency independent complex gains, 
was performed individually for each of the coarse channels using a 
10-second interval where the Sun was in a relatively quiescent state.  
These gain solutions were then applied to the full time interval.
No absolute flux calibration was possible, owing to the lack of 
observations of a suitable calibrator.
The amplitude part of the bandpass was calibrated using the total power
spectra from the same quiescent time interval as used for
self-calibration. 
As will become evident later, the choice of a spectral index only impacts
the underlying spectral slope.
We assumed a spectral index $\alpha$=2.6 for the
quiet Sun ($S_{\nu}\propto \nu^{\alpha}$ where $S_{\nu}$ is the
flux density and $\nu$ is the observing frequency);
this choice is based on 32T observations during the recent deep 
solar minimum (2008 November), which yielded 
$\alpha=2.6\pm$0.4, a value consistent with \citet{Erickson77}.
All the data presented here correspond to the east-west (XX) polarization.
The wide FoV MWA tiles are expected to have significant (but stable)
cross-polarization leakage, and an absolute polarization calibration 
was not attempted.

Imaging and deconvolution were performed using the standard CLEAN
algorithm with robust (${\cal R}$=0) weighting, resulting in a synthesized
beam at the band center of $961''\times796''$ (FWHM). 
For the images presented here, the data were averaged over a single
coarse channel during the gridding process and restored with a 
circular beam with FWHM $800''$. 
The edges of the coarse channels could not be calibrated satisfactorily, 
hence the first six and the last four spectral channels of each coarse 
channel were flagged during imaging.

%%%%%%%%%%%%%%%%%%%%%%%%%%%%%%%%%%%%%%%%%%%%%%%%%%%%%%%%%%%%%%%%%%%%%%%%%%%
\begin{landscape}
\begin{figure}
\hbox{
% Landscape mode 2x2 (DONE)
\resizebox{0.5\hsize}{!}{\includegraphics[trim=0mm 0mm 0mm 0mm, clip=true]{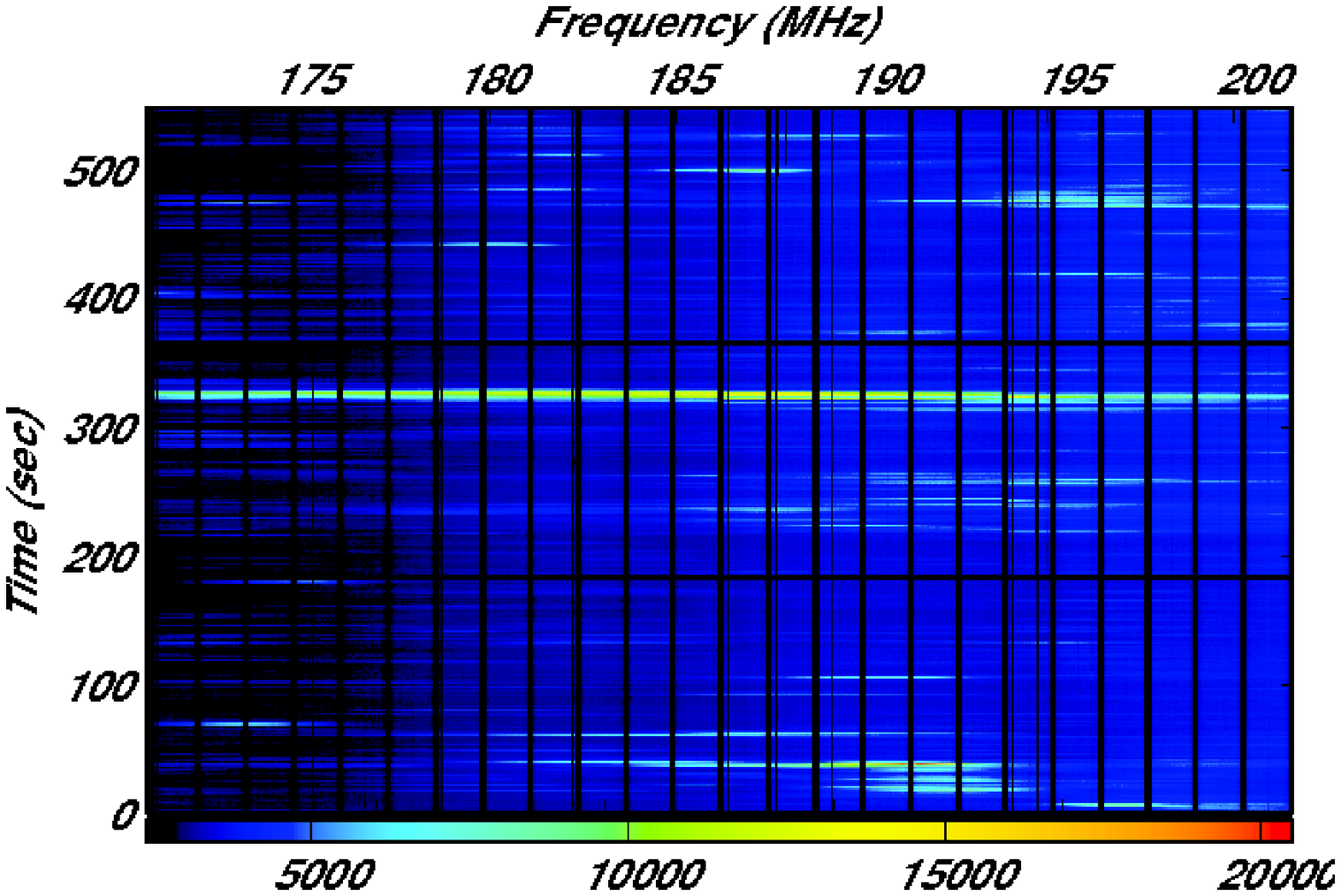}}
\resizebox{0.5\hsize}{!}{\includegraphics[trim=0mm 0mm 0mm 0mm, clip=true]{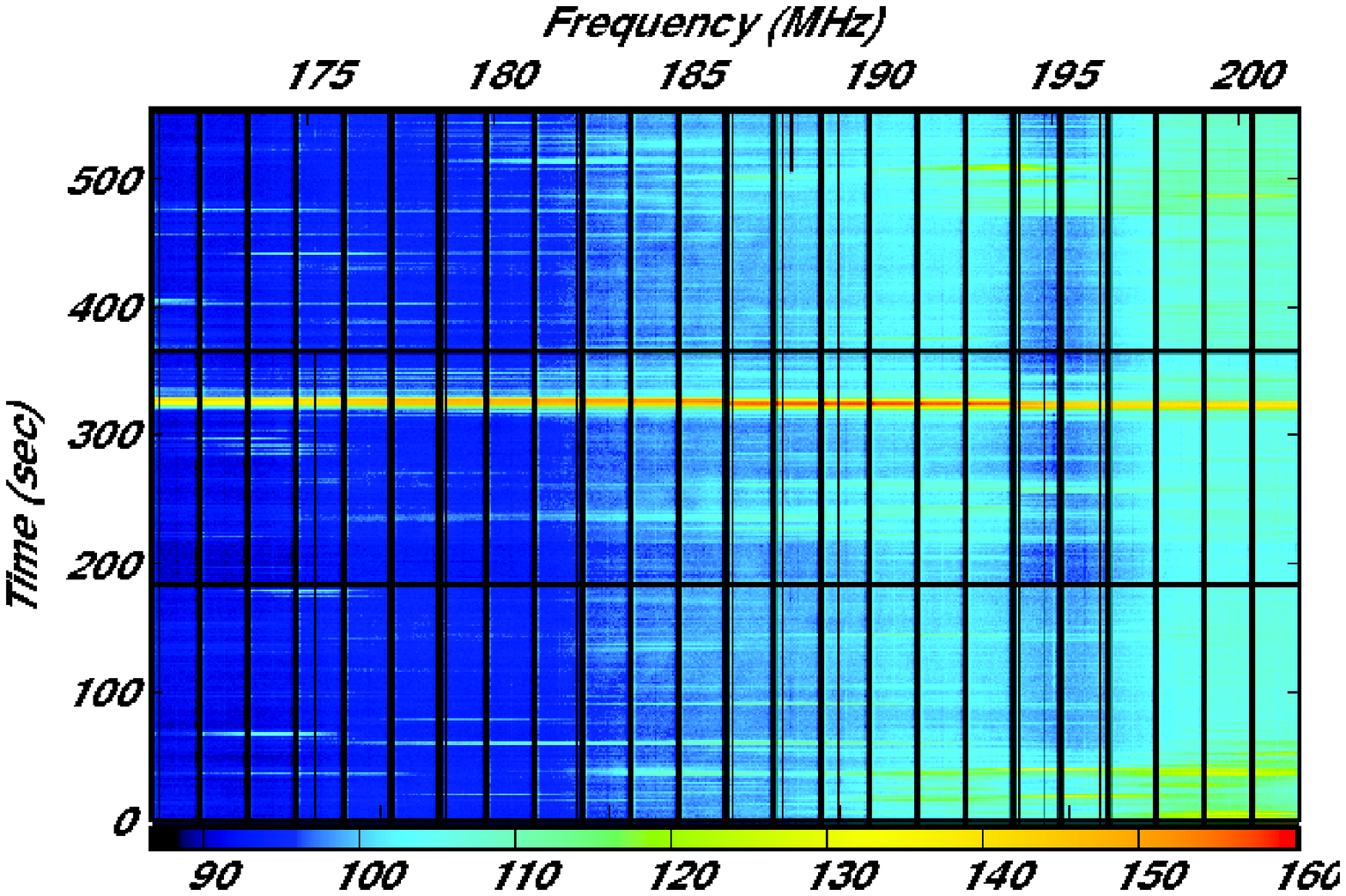}}
%\resizebox{0.5\hsize}{!}{\includegraphics[trim=0mm 0mm 0mm 0mm, clip=true]{FigAnew.eps}}
%\resizebox{0.5\hsize}{!}{\includegraphics[trim=0mm 0mm 0mm 0mm, clip=true]{FigBnew.eps}}
}
\hbox{
% Landscape mode 2x2
\resizebox{0.495\hsize}{!}{\includegraphics[trim=0mm 0mm 0mm 0mm, clip=true]{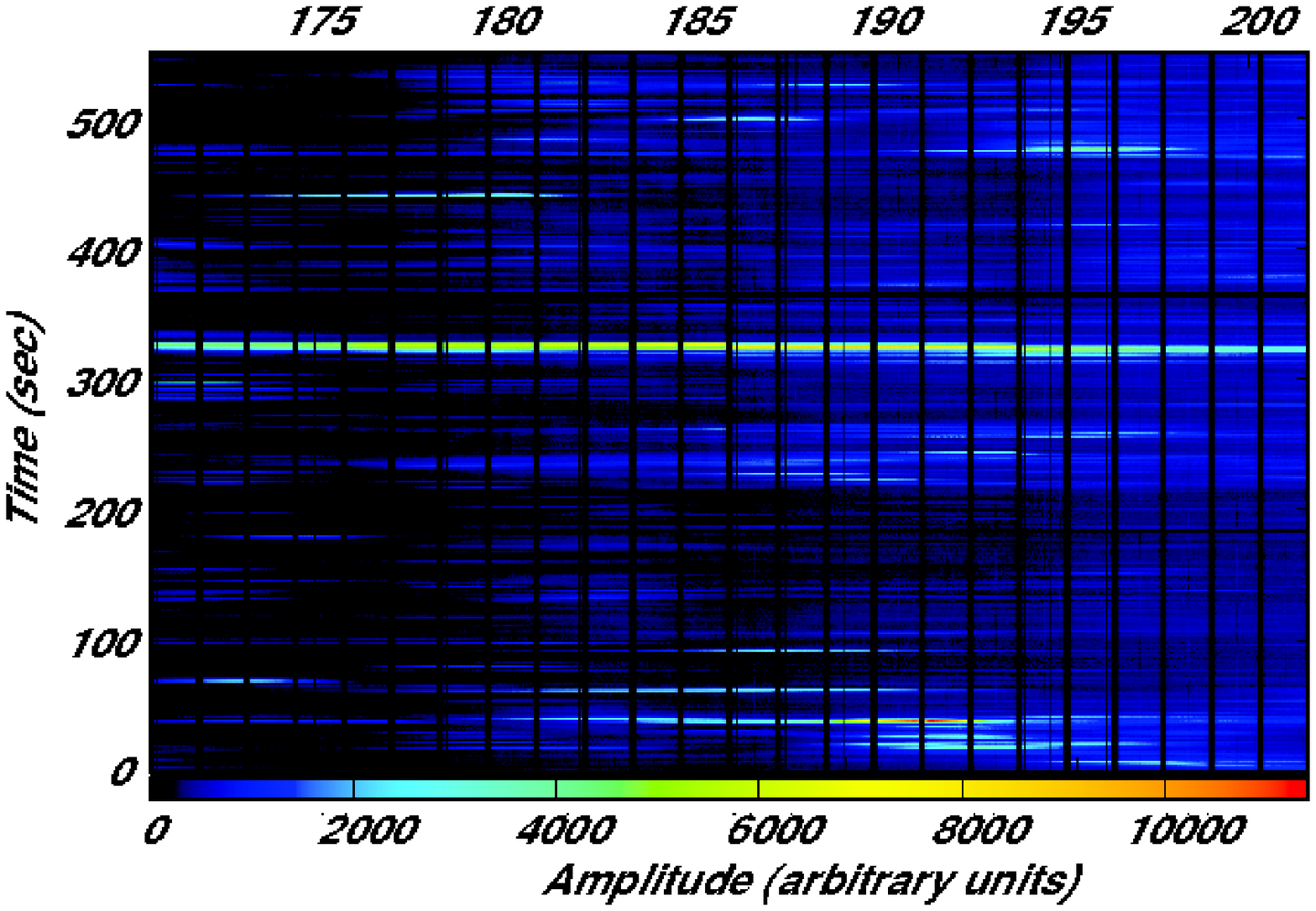}}
\resizebox{0.495\hsize}{!}{\includegraphics[trim=0mm 0mm 0mm 0mm, clip=true]{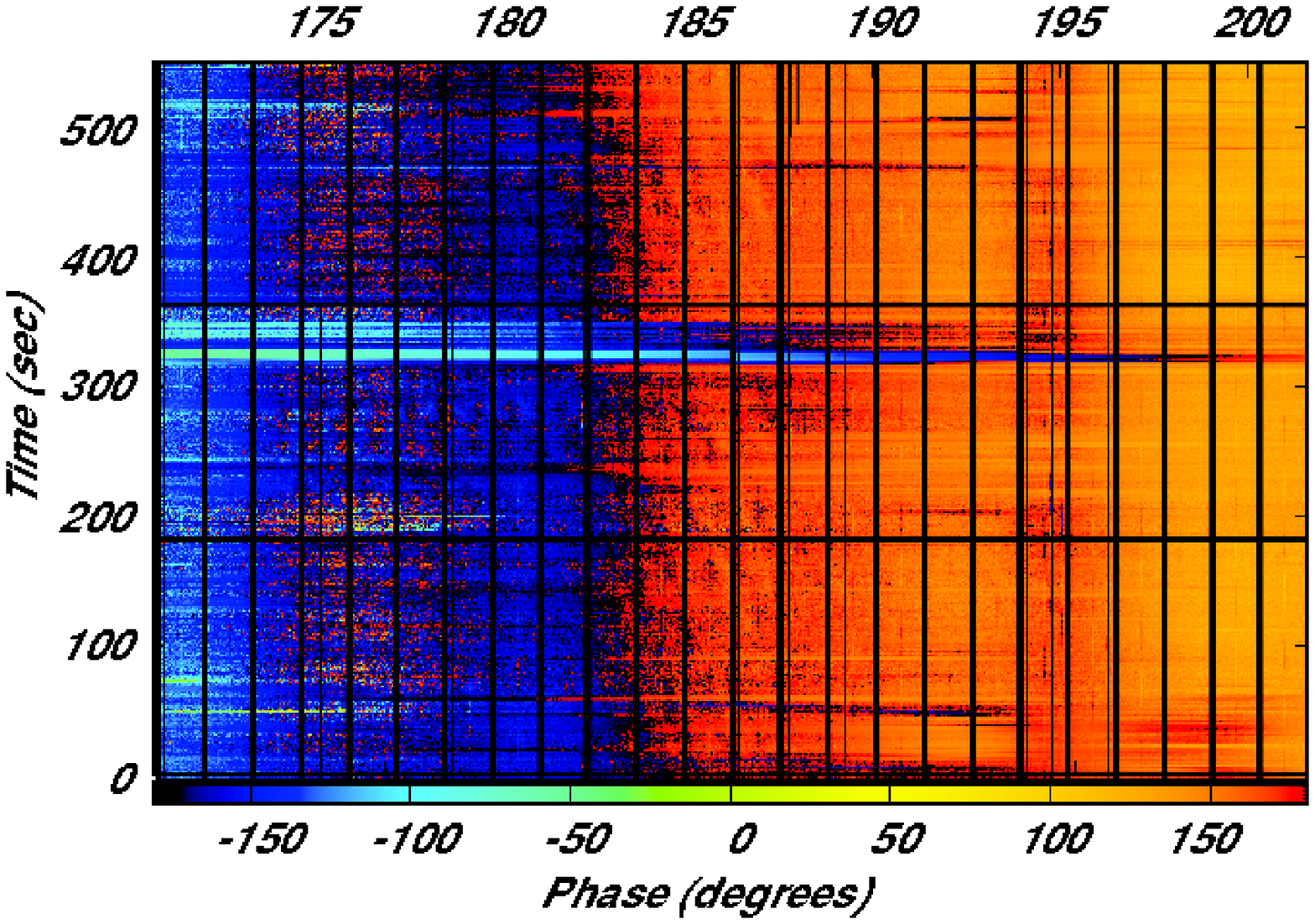}}
%\resizebox{0.495\hsize}{!}{\includegraphics[trim=0mm 0mm 0mm 0mm, clip=true]{FigCnew.eps}}
%\resizebox{0.495\hsize}{!}{\includegraphics[trim=0mm 0mm 0mm 0mm, clip=true]{FigDnew.eps}}
}
\caption{ 
Observed visibilities as a function of frequency (XX polarization).
The left panels show amplitude (arbitrary units), and
  the right panels phase (in degrees).
The top and bottom rows correspond to baselines with projected
  lengths of $\sim$40$\lambda$ and $\sim$196$\lambda$, respectively, at 
the band center (186.26~MHz). 
The frequency ({\em x}-axis) ranges from 170.8 to 201.6~MHz; the time ({\em y}-axis) spans 550~s starting at 04:24:53~UT on 2010 March 27.
The horizontal bar below each panel shows the color coded scale and the dark vertical streaks arise from flagging of the coarse channel edges (Sec.~\ref{Sec:data}).
}
\label{Fig:visibility_amp_phase}
\end{figure}
\end{landscape}
%%%%%%%%%%%%%%%%%%%%%%%%%%%%%%%%%%%%%%%%%%%%%%%%%%%%%%%%%%%%%%%%%%%%%%%%%%%%%

\section{Analysis and Results}
\label{Sec:analysis}
Amplitudes and phases observed on a representative short ($\sim$64~m)
and long ($\sim$315~m) projected baseline are shown in 
Fig.~\ref{Fig:visibility_amp_phase}. 
Baselines of different lengths and orientations measure independent
Fourier components of the source structure. 
The examples shown here illustrate several key features of the data. 
Only one polarization is shown as both polarizations show very similar behavior.
The most prominent feature in Fig.~\ref{Fig:visibility_amp_phase},
visually obvious on all baselines in both amplitude and phase, 
spans the entire observing band around 04:30:10~UT.
We hereafter refer to this as the ``broadband'' feature.
Its properties are similar to a weak type~III burst (Sec.~\ref{Sec:discussion}).
In addition, a large number of shorter-lived, narrowband 
features are evident in the visibility amplitudes. 
Many of these amplitude variations are accompanied by corresponding 
variations in phase, implying a change in the brightness temperature
($T_B$) distribution in the corona.
These features become more numerous and prominent visually with increasing
baseline length and are referred to as ``narrowband'' features 
hereafter.
%Again, the magnitude of phase variations and the number of features
%with associated phase variations increase with baseline length.
A dearth of truly ``quiescent" periods is also evident, and 
the ubiquitous modest variations seen in the spectral structure 
are solar, not instrumental, in origin.
%Though difficult to see in this representation, there is a shift of
%1.0 s (2.0 s) between the first half and the third (fourth) quarter
%of the band, for a part of these data {\bf (XX:XX:XX - XX:XX:XX UT
%don't know if this is easy or useful to provide)} due to an
%instrumental glitch.

\begin{figure}
{
\resizebox{1.0\hsize}{!}{\includegraphics[trim=0mm 15mm 5mm 0mm, clip=true]{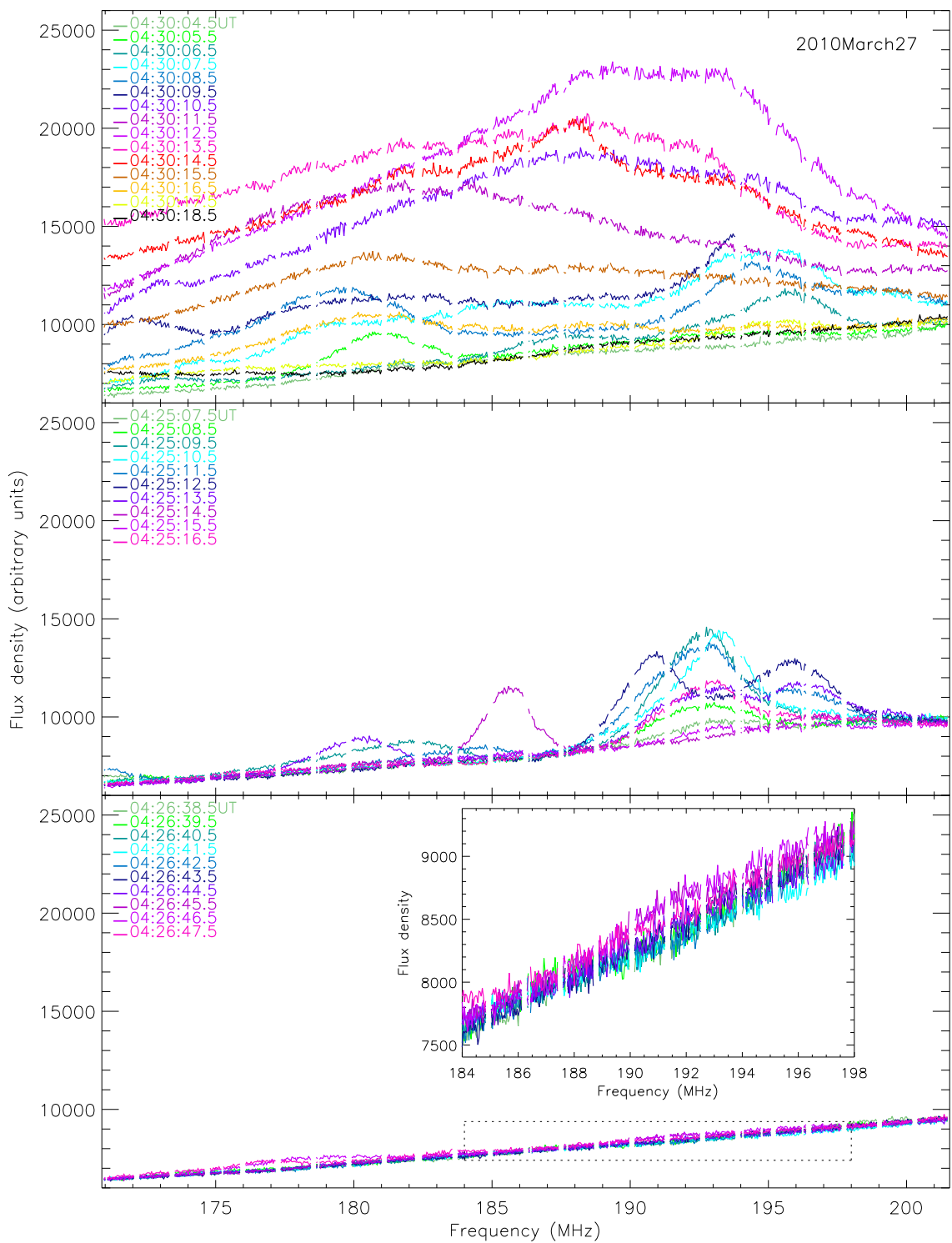}}
}
\caption{
Sample autocorrelation spectra from tile 6, X polarization. 
{\em Top panel:} a 14~s period bracketing the most 
prominent ``broadband'' feature seen in Fig.~\ref{Fig:visibility_amp_phase}.
{\em Middle panel:} variations seen across 10~s for one of 
the numerous shorter-lived ``narrowband'' intensity enhancements. 
{\em Bottom panel:} a 10 second interval exhibiting little intensity 
variation. 
The inset in the bottom panel zooms in on a part of the band.
%The top panel data have been corrected for the instrumental glitch mentioned in context of Fig \ref{Fig:visibility_amp_phase}.
As with Fig. \ref{Fig:visibility_amp_phase}, the data gaps reflect flagged data.
}
\label{Fig:spectral_cuts}
\end{figure}

To further illustrate the diversity in the types of 
emission seen in the short span of data presented here, as well as
its spectrally complex and highly time-variable behavior, 
Fig. \ref{Fig:spectral_cuts} shows several series of autocorrelation 
(total power) spectra.
%%%%%%%%%%%%%%%%%%%%%%%%%%%%%%%%%%%%%%%%%%%%%%%%%%%%%%%%%%%%%%%%%%%%%%
%%%%%%%%%%%%%%%%%%%%%%%%%%%%%%%%%%%%%%%%%%%%%%%%%%%%%%%%%%%%%%%%%%%%%%%%
The bottom panel shows one of the few quiescent intervals; the baseline 
spectral slope seen here is due to the spectral index of the quiet
Sun, dominated by thermal emission. 
An amplitude variation of $\sim$5\% over a 10 s interval is seen even
in ``featureless'' parts of the data.
The middle panel shows a series of spectra illustrating one of the
numerous ``narrowband'' features seen in
Fig. \ref{Fig:visibility_amp_phase}. 
These features show remarkably rapid evolution in spectral shape and
intensity. They typically span $\sim$5--10~MHz in bandwidth, 
outside of which the spectral flux density returns to that of the quiet Sun.
Their peak flux density approaches $\sim$1.5 times the quiescent solar flux density at 
that frequency.  Finally, the top panel of Fig.~\ref{Fig:spectral_cuts} 
shows the dynamic behavior of the ``broadband'' feature around 04:30:10~UT. 
Its bandwidth exceeds the $\sim$30~MHz observing bandwidth, and at its peak,
its flux density is $\sim$2.5 times the quiescent solar flux density.
The abrupt changes from one second to the next imply that the physical 
changes leading to the production of this emission are temporally 
undersampled. 

\begin{figure}
{
\resizebox{0.324\hsize}{!}{\includegraphics[angle=0, trim=0mm 0mm 0mm 0mm, clip=true]{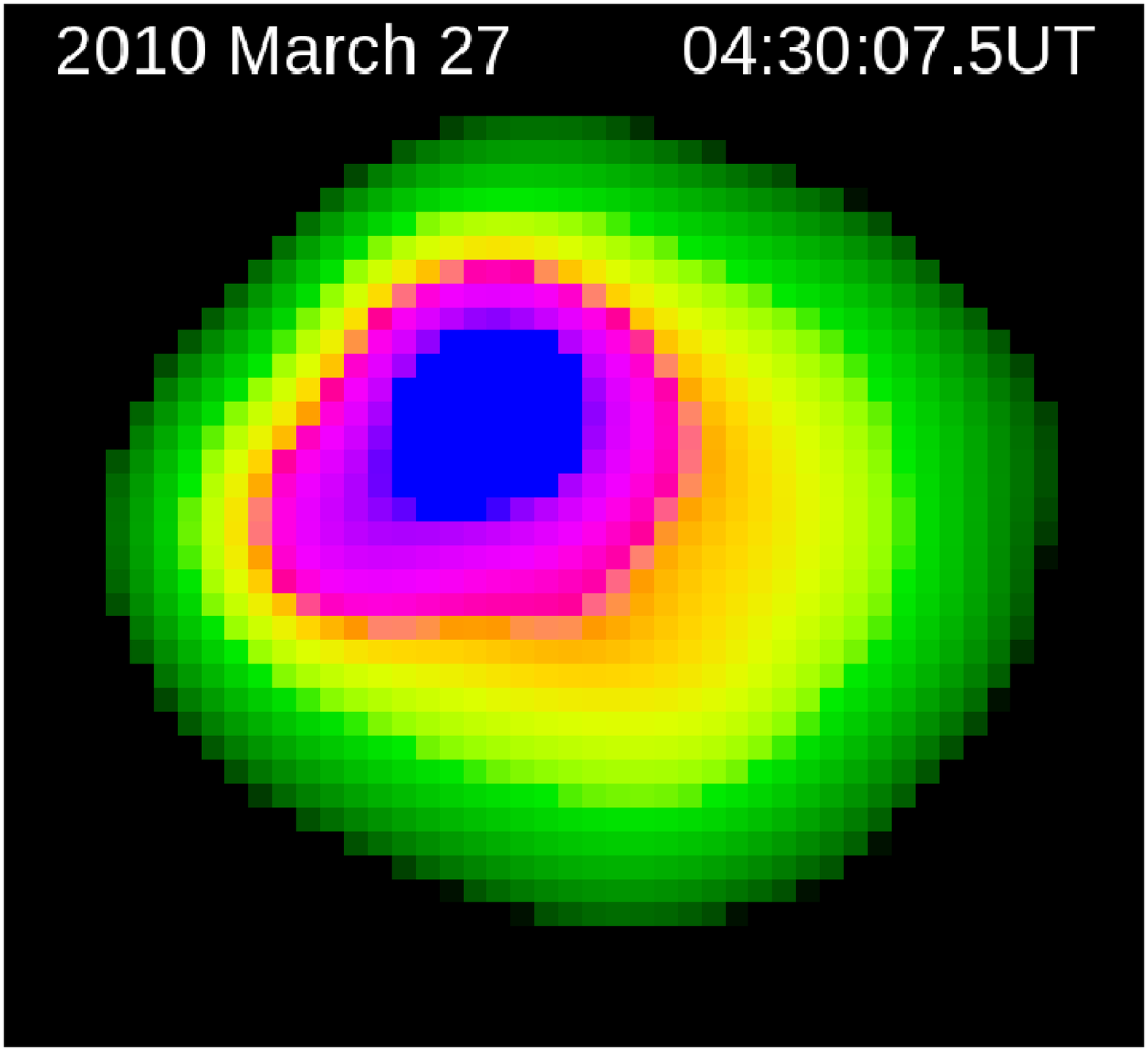}}
\resizebox{0.324\hsize}{!}{\includegraphics[angle=0, trim=0mm 0mm 0mm 0mm, clip=true]{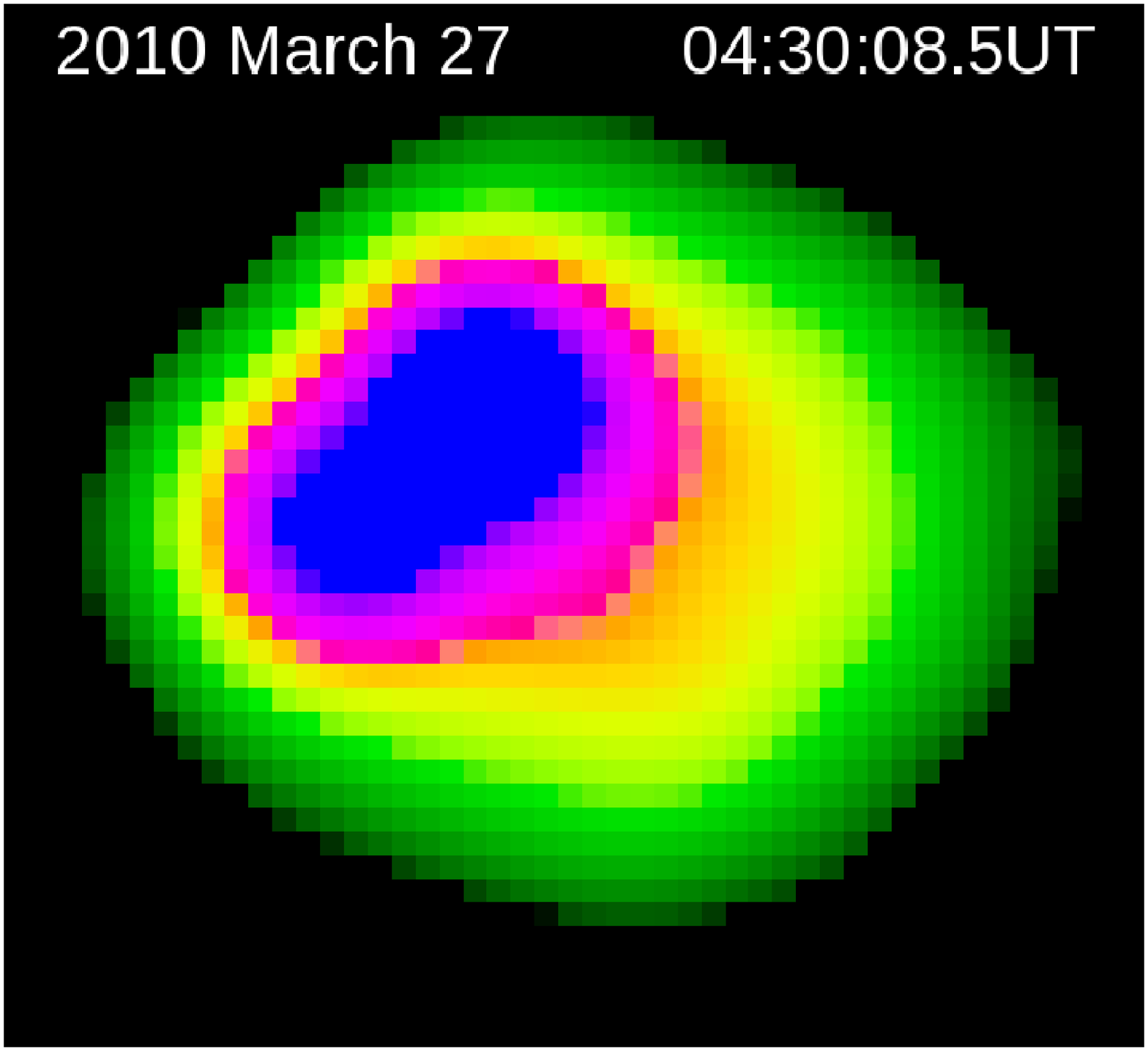}}
\resizebox{0.324\hsize}{!}{\includegraphics[angle=0, trim=0mm 0mm 0mm 0mm, clip=true]{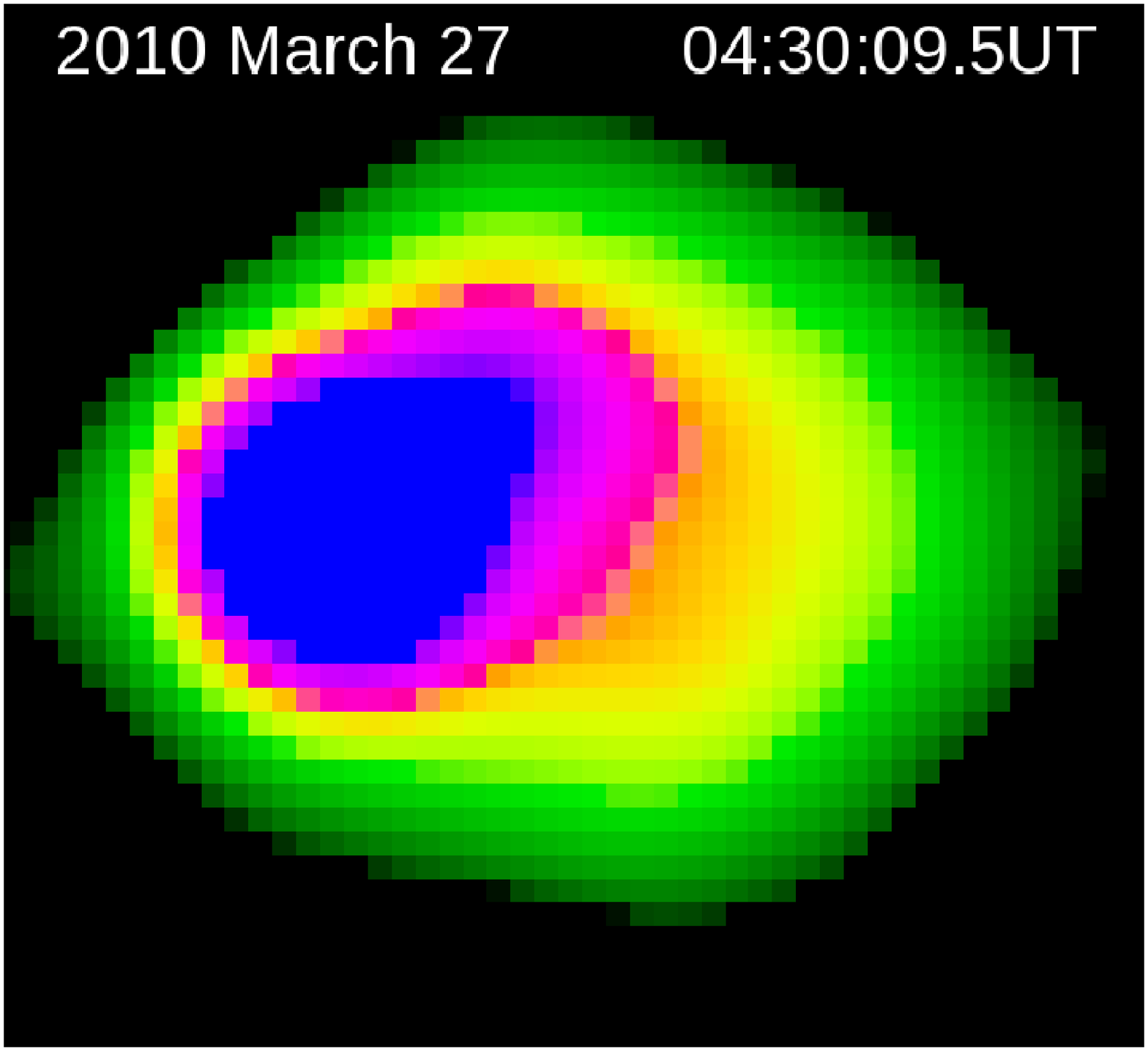}}

\resizebox{0.324\hsize}{!}{\includegraphics[angle=0, trim=0mm 0mm 0mm 0mm, clip=true]{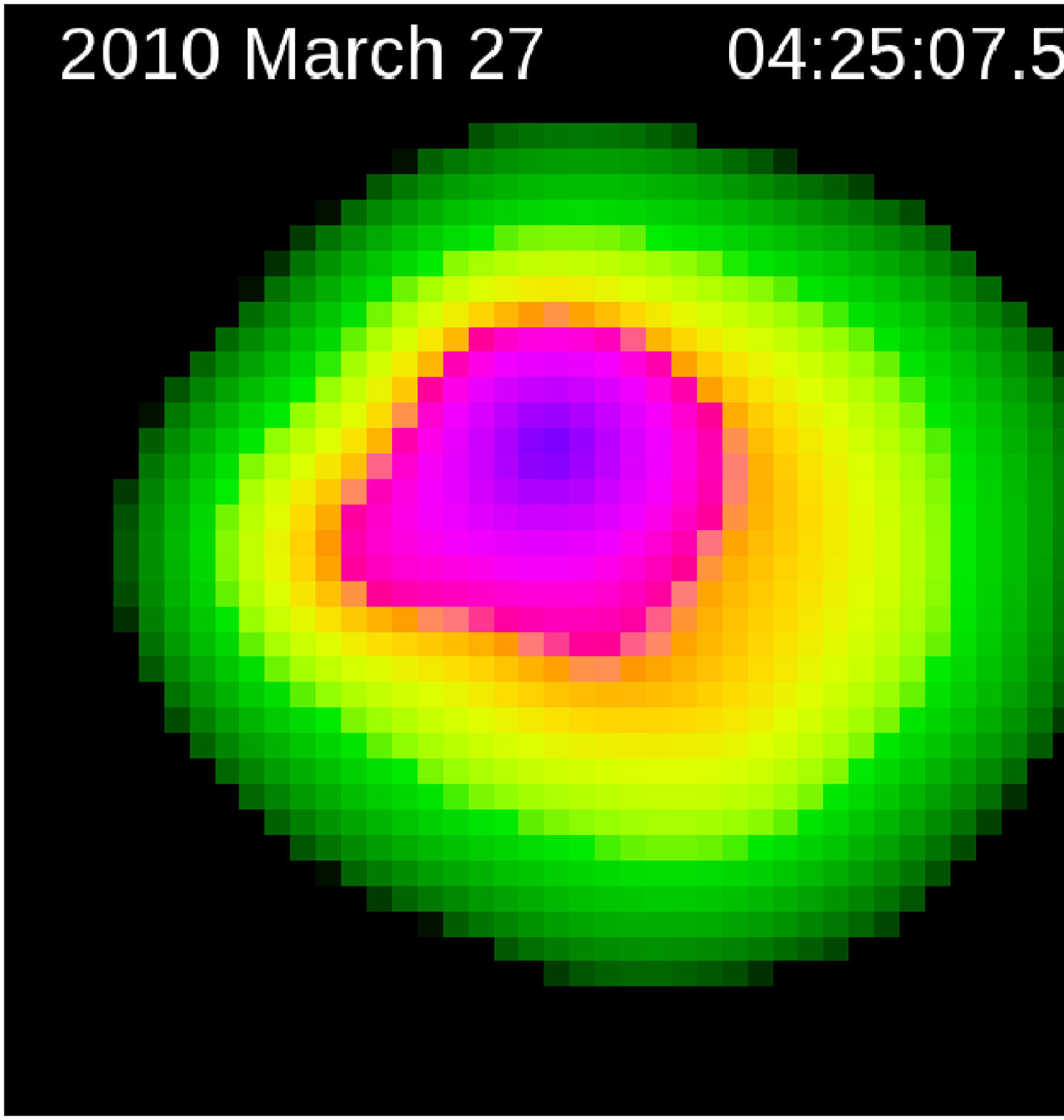}}
\resizebox{0.324\hsize}{!}{\includegraphics[angle=0, trim=0mm 0mm 0mm 0mm, clip=true]{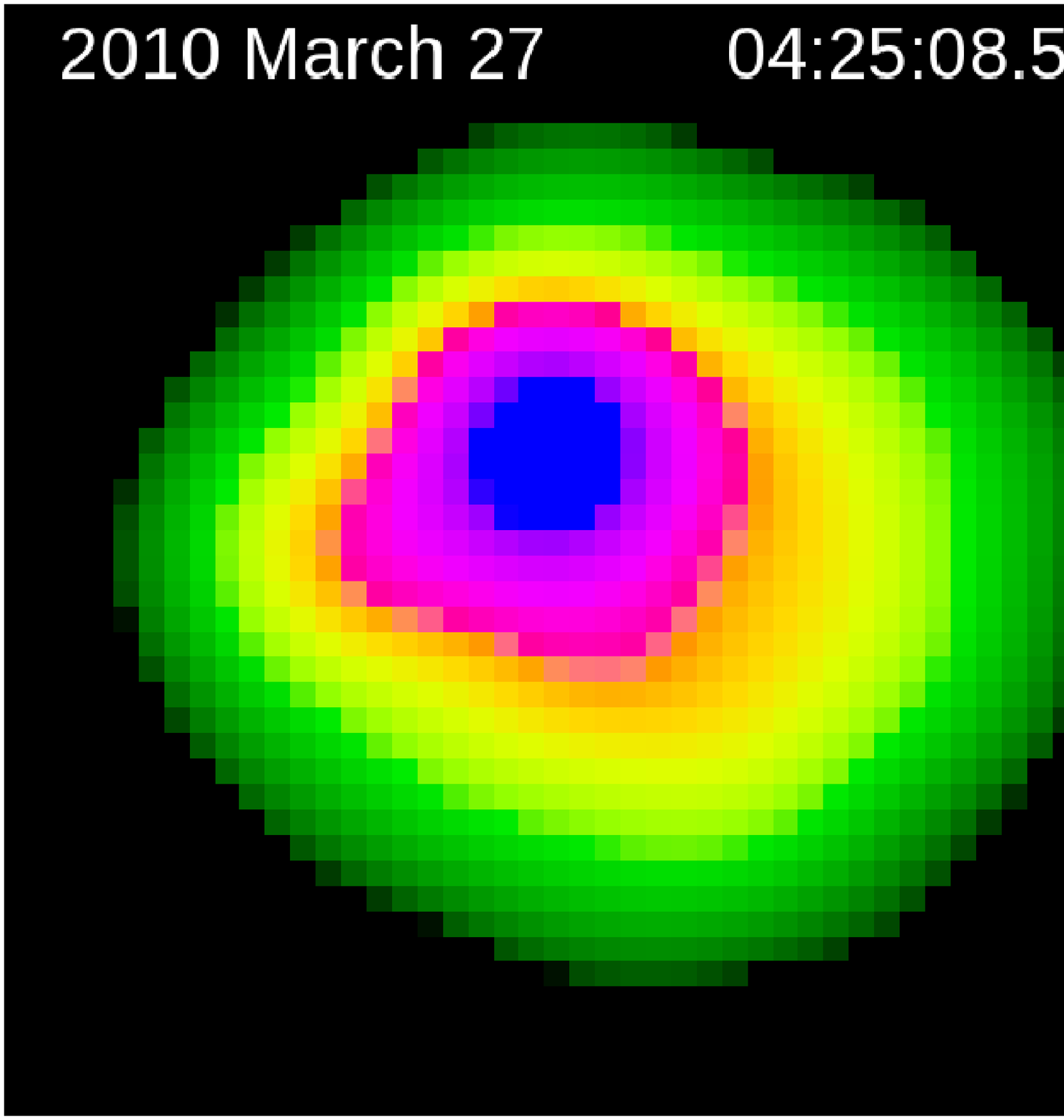}}
\resizebox{0.324\hsize}{!}{\includegraphics[angle=0, trim=0mm 0mm 0mm 0mm, clip=true]{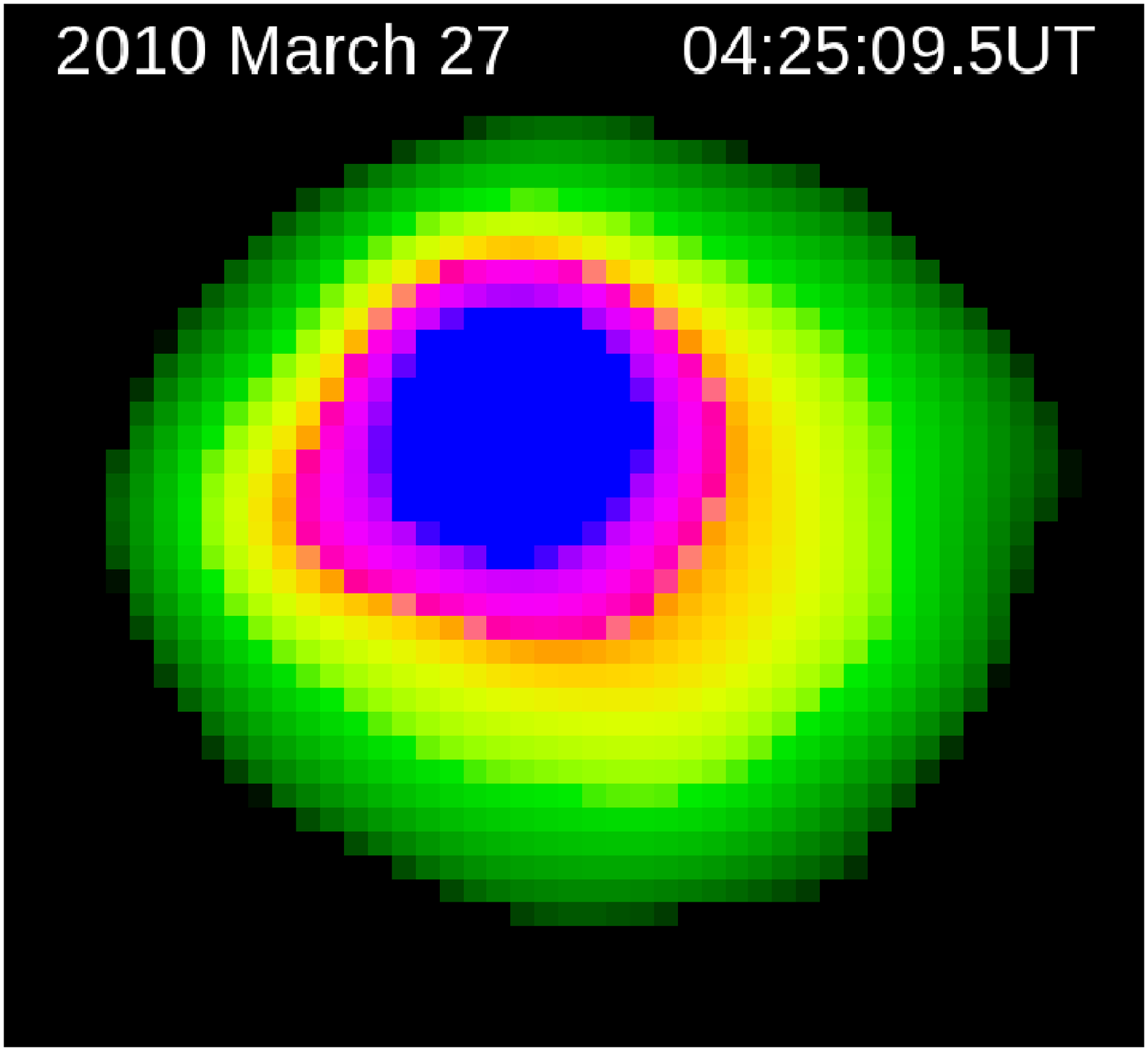}}

\hspace{0.23\hsize}
\resizebox{0.325\hsize}{!}{\includegraphics[angle=0, trim=0mm 0mm 0mm 0mm, clip=true]{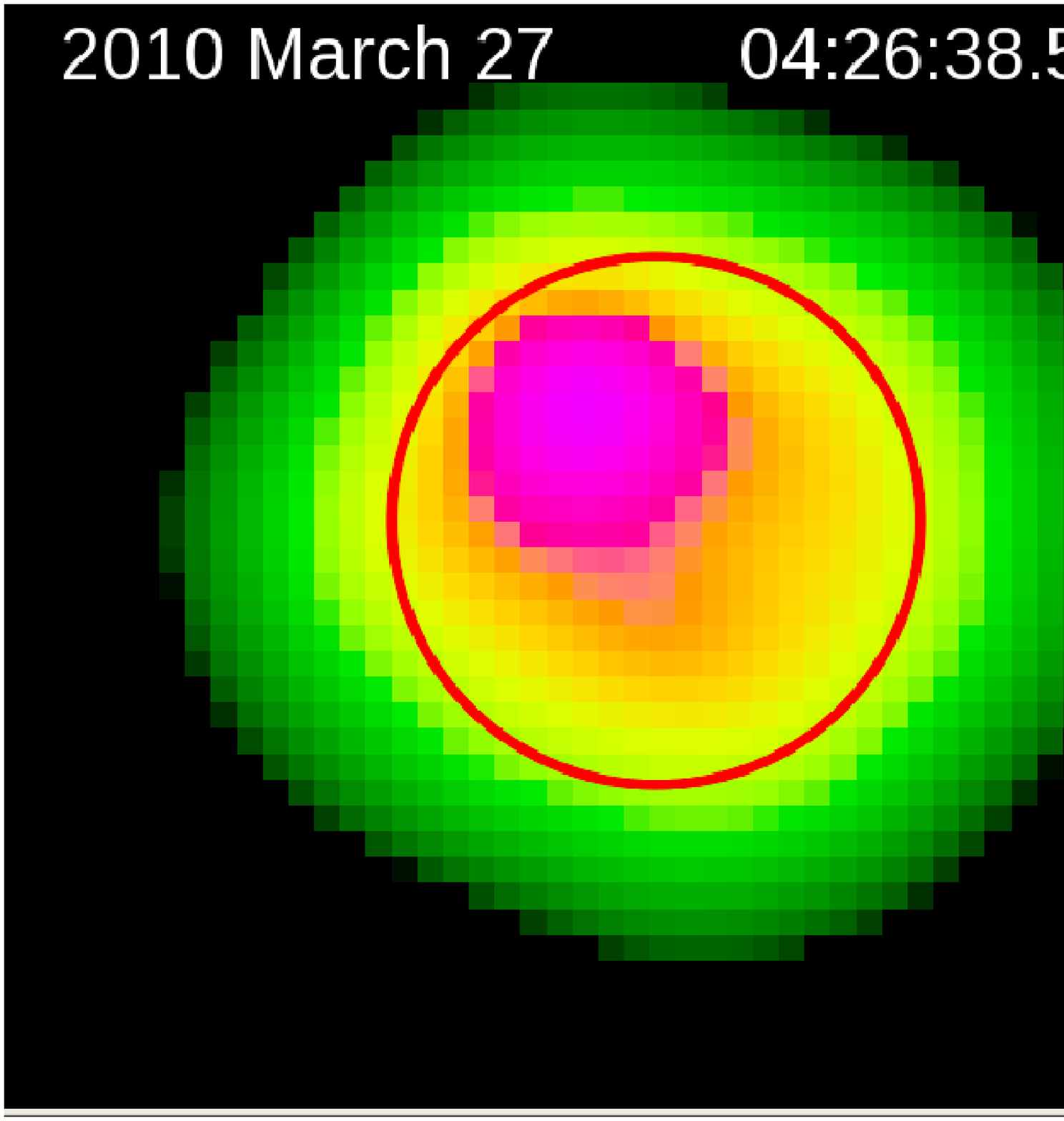}}
\put(0,24.0) {
\resizebox{0.188\hsize}{!}{\includegraphics[angle=0, trim=0mm 0mm 0mm 0mm, clip=true]{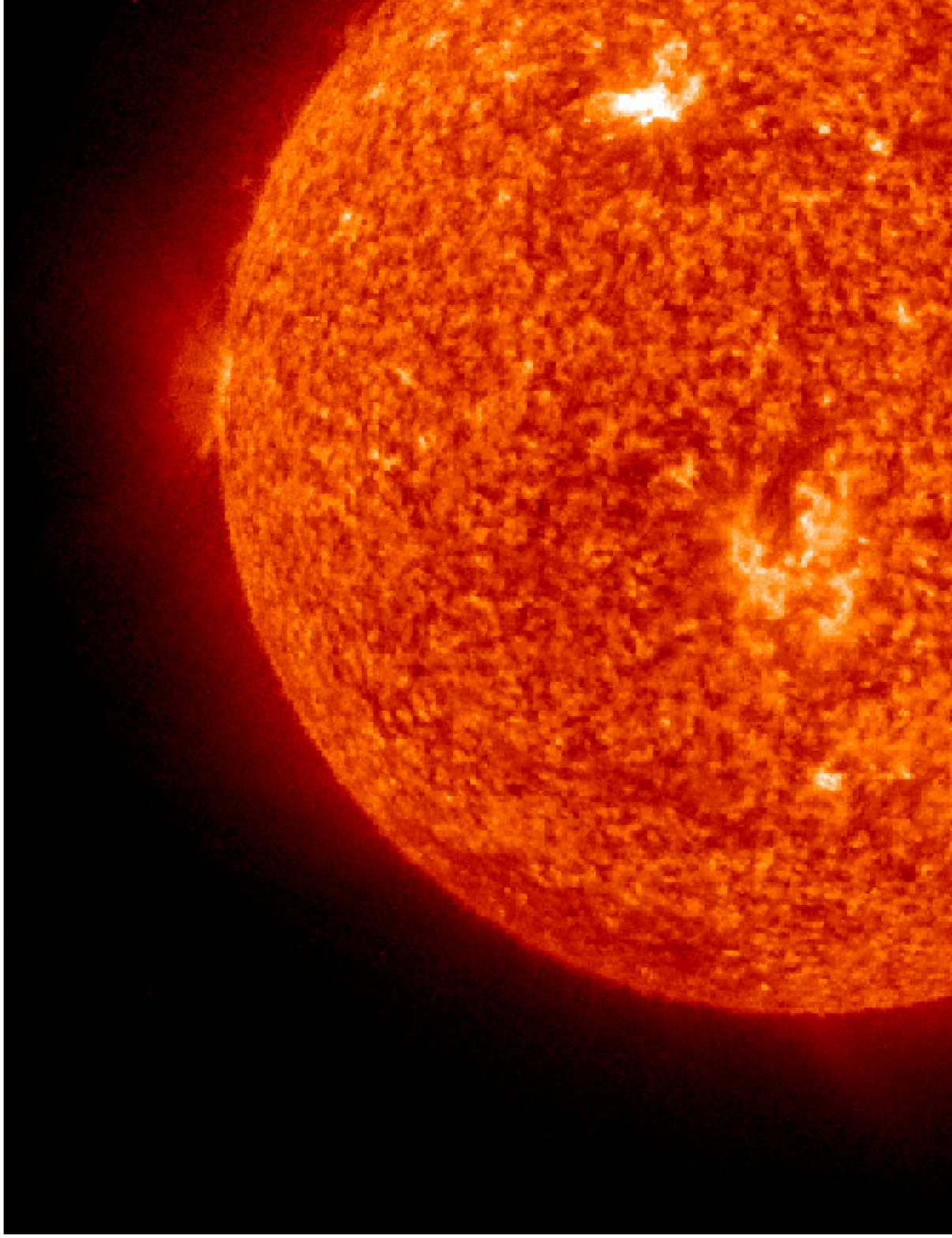}}
}
}
\caption{
{\em Top and middle rows:} sets of images, 1~s apart, 
from the ``broadband'' and one of the numerous ``narrowband''
features, respectively, highlighting the rapid temporal variation in the 
emission morphology. 
Animations including these panels are available in the on-line journal.
{\em Bottom row:} A 32T radio image of the quiescent Sun (left)
and a 304{\rm \AA} image from SOHO/EIT (right), taken a few hours
earlier (01:19~UT).
The bright region in the northeastern quadrant is the region 11057.
The red circle on the 32T image represents the size of the optical solar 
disc and has been registered by aligning the centroid of radio emission 
at the location of the active region 11057 with its counterpart on the
304{\rm \AA} image, ignoring the time offset.
All radio images are at 193.3~MHz (XX polarization) and have
the same intensity scale (0--2500, arbitrary units). 
All images have celestial north on top.
}
\label{Fig:movie_in_time}
\end{figure}

The data presented here permit a high fidelity image of the Sun for
every individual time and frequency slice (i.e. every pixel on the
dynamic spectra shown in Fig. \ref{Fig:visibility_amp_phase}). 
Fig. \ref{Fig:movie_in_time} shows a collage of images that
illustrates the rapid evolution of the appearance of the solar corona
during our observations. 
%%%%%%%%%%%%%%%%%%%%%%%%%%%%%%%%%%%%%%%%%%%%%%%%%%%%%%%%%%%%%%%%%%%%%%%%
%%%%%%%%%%%%%%%%%%%%%%%%%%%%%%%%%%%%%%%%%%%%%%%%%%%%%%%%%%%%%%%%%%%%%%%%%%%%%
Although the spatial resolution of the 32T is limited, the fidelity
and dynamic range of these images is unprecedented at these
frequencies; the dynamic range of $\sim$2500 exceeds that of earlier images by about 
an order of magnitude \citep[cf.][]{Mercier06, Mercier09}.
The higher fidelity and dynamic range enables us not only to image 
the quiescent solar emission in the presence of much brighter features, 
unlike most earlier observations \citep[e.g.,][]{Kai70, Kundu86,
Krucker95, Vilmer02, Ramesh_etal2010}, but also 
to track low level variations in the coronal emission.
The bottom panels in Fig.~\ref{Fig:movie_in_time} illustrate 
the morphological relationship between our 32T radio images and 
the extreme ultraviolet (EUV) images taken on the same day. 
A persistent feature coinciding with the  location of the NOAA 
Space Weather Prediction Center (SWPC) region 11057 (hereafter region 
11057), is seen in the solar radio images, even at quiescent times.

Rapid and significant changes in the radio morphology of the Sun on
timescales of seconds are clearly visible in the images presented in
Fig.~\ref{Fig:movie_in_time} and are further highlighted in the two ``movies''
available in the on-line journal. 
A significant fraction of this variability maps back to the location of
the region 11057.
Subtle changes in $T_B$ as a function of time are also discernible at
the locations of the active regions on the Eastern limb and close to the 
center of the disc.
%An apparent motion of the bright emission feature associated with region 11057 can be seen in the animation.

\begin{figure}
\begin{center}
{
\resizebox{0.33\hsize}{!}{\includegraphics[angle=0, trim= 15mm 15mm 15mm 0mm, clip=true]{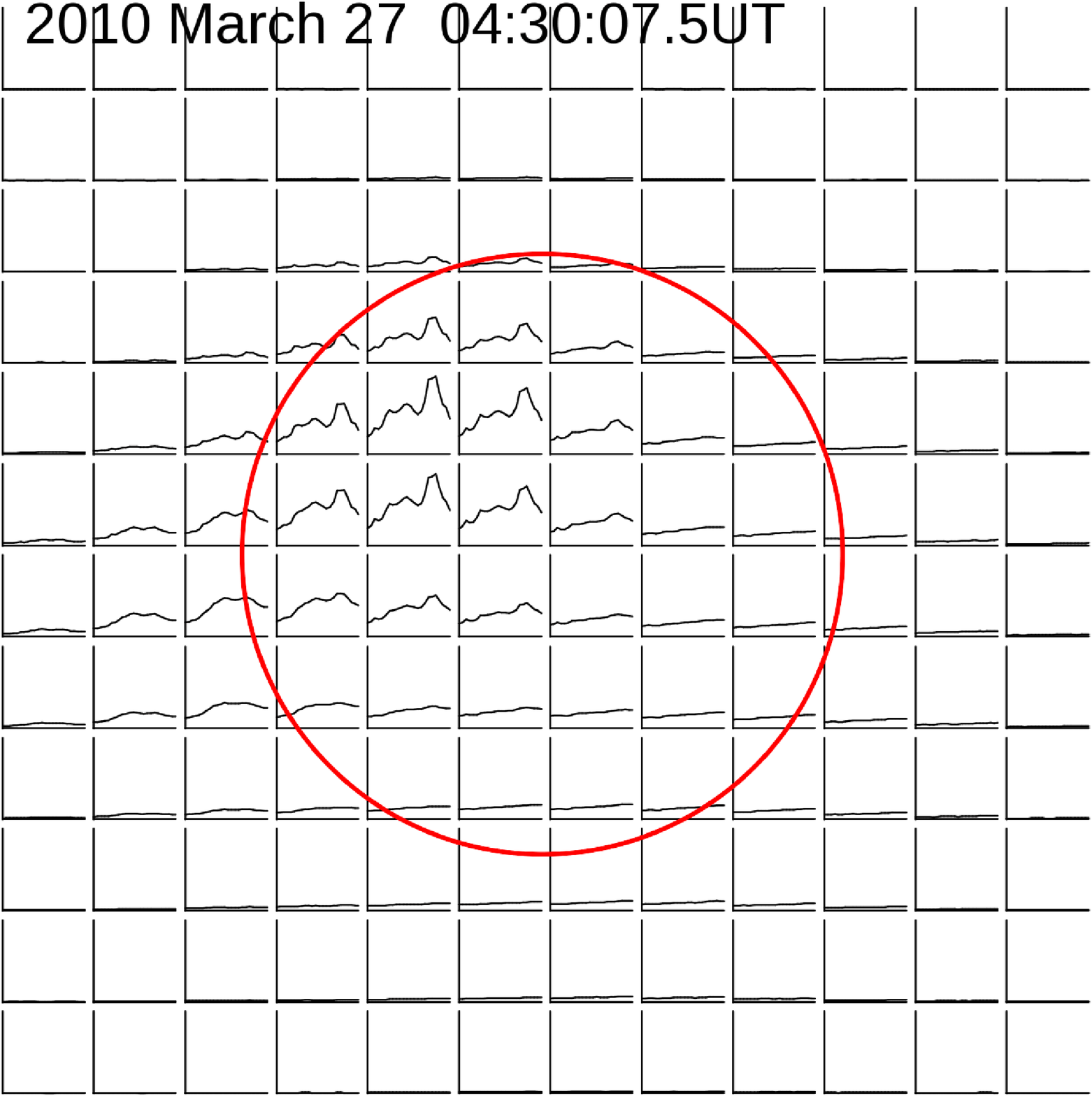}}
\resizebox{0.33\hsize}{!}{\includegraphics[angle=0, trim= 15mm 15mm 15mm 0mm, clip=true]{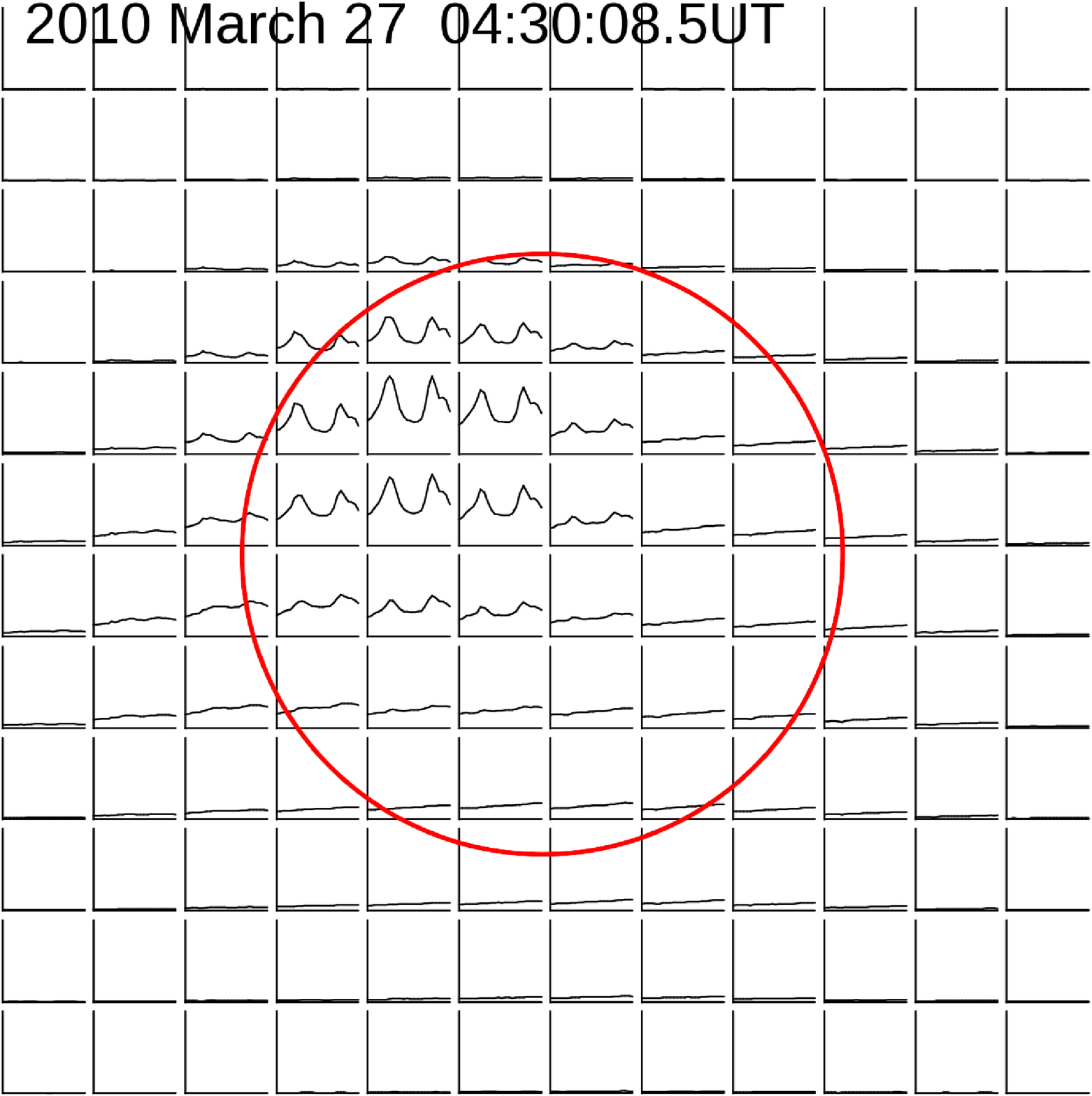}}

\resizebox{0.33\hsize}{!}{\includegraphics[angle=0, trim= 15mm 15mm 15mm 0mm, clip=true]{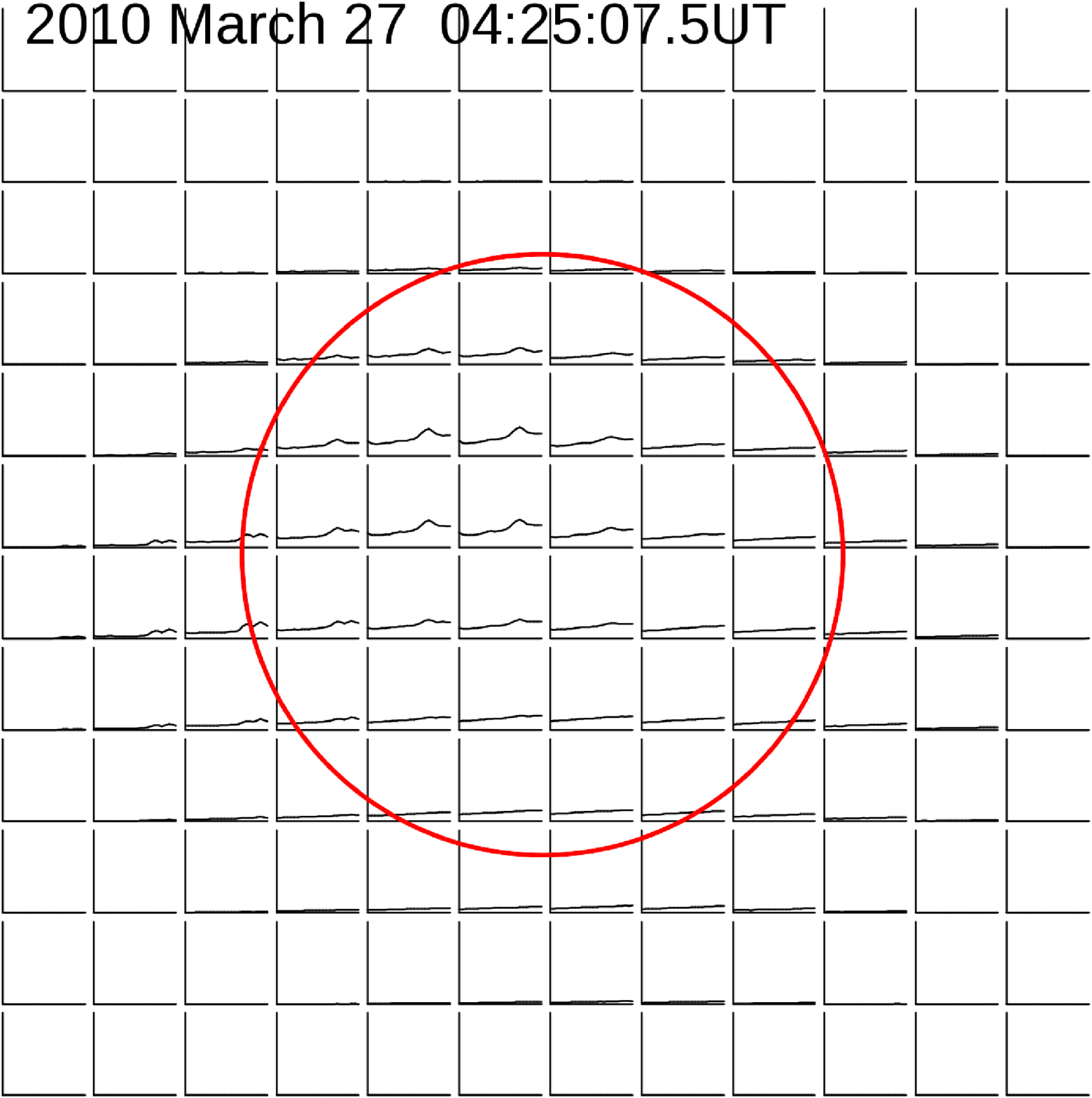}}
\resizebox{0.33\hsize}{!}{\includegraphics[angle=0, trim= 15mm 15mm 15mm 0mm, clip=true]{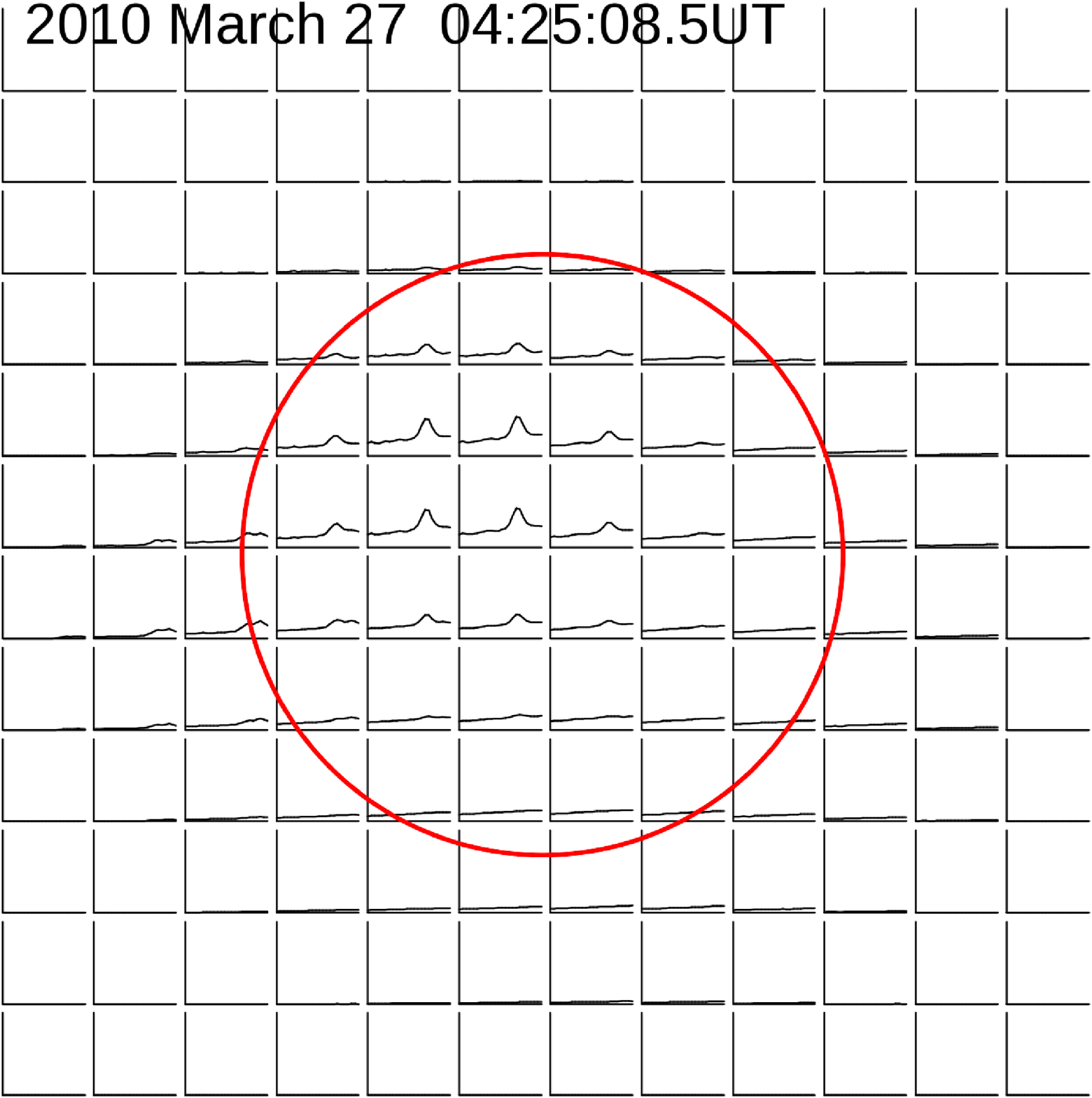}}

\resizebox{0.33\hsize}{!}{\includegraphics[angle=0, trim= 15mm 15mm 15mm 0mm, clip=true]{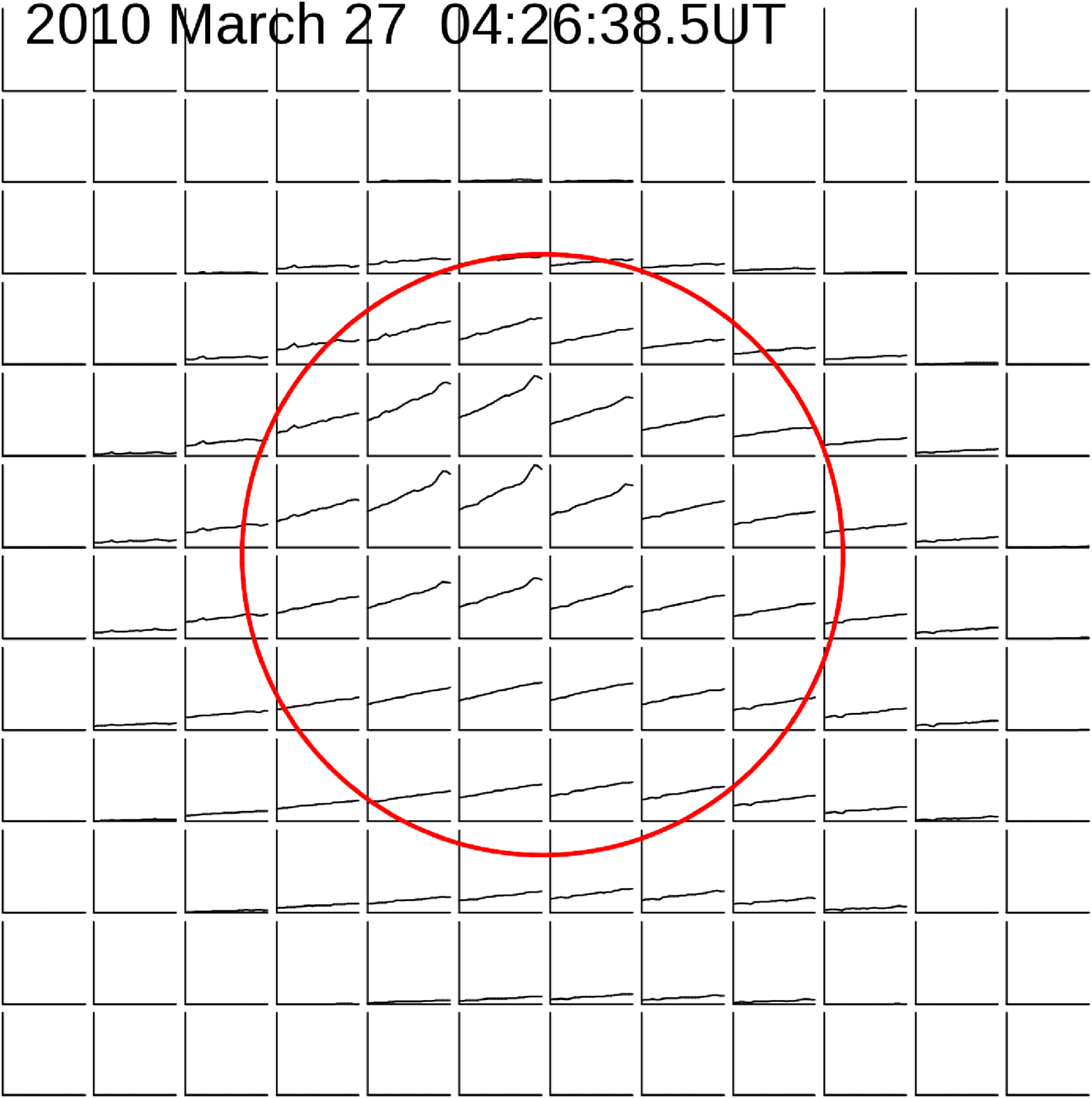}}
\resizebox{0.33\hsize}{!}{\includegraphics[angle=0, trim= 15mm 15mm 15mm 0mm, clip=true]{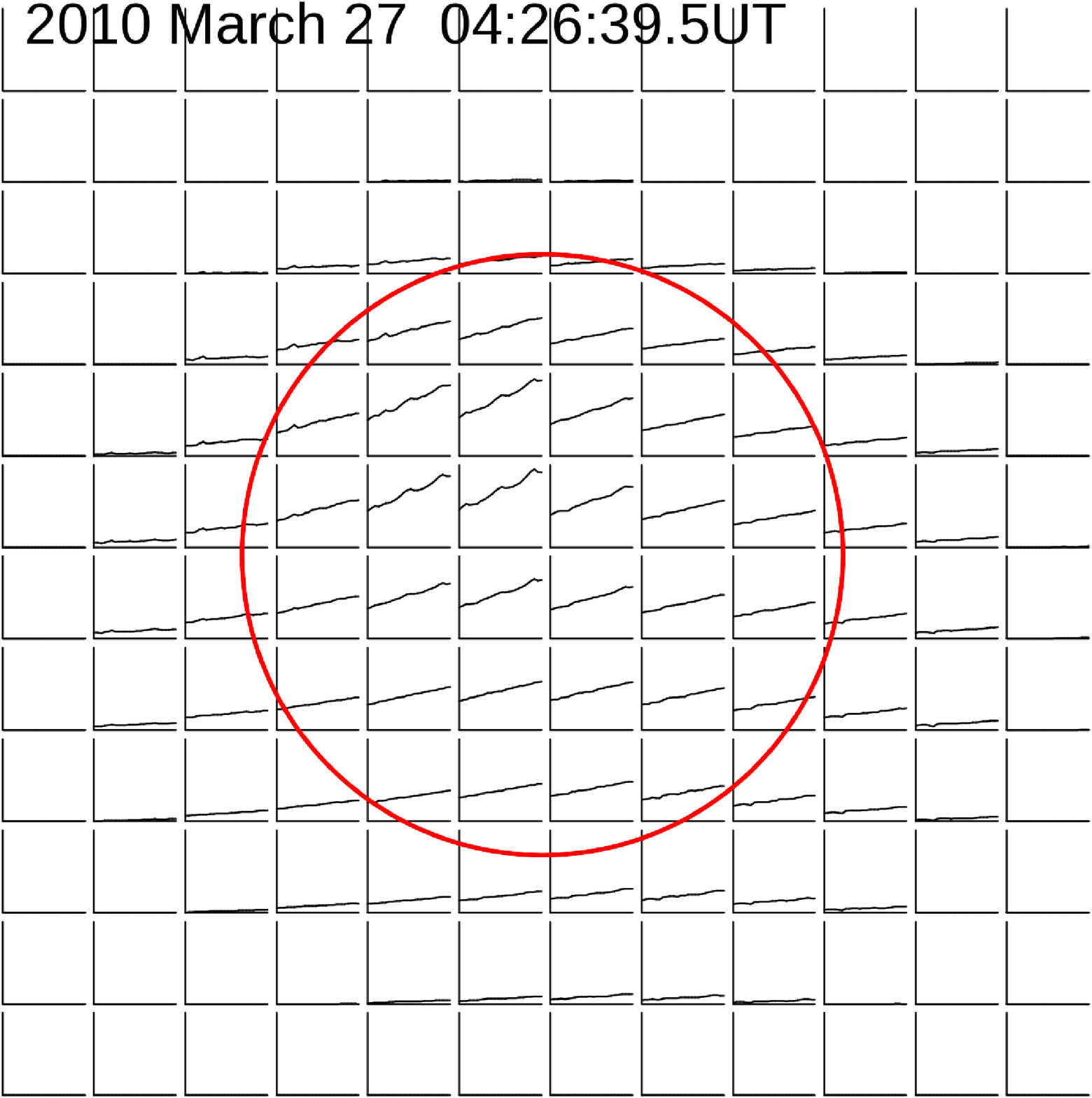}}
}
\end{center}
\caption{ 
Spatially localized spectra of emission across the solar disk
at several of the same times shown in Fig. \ref{Fig:movie_in_time}.
Celestial north is on top, and the red circle marks the size of 
the optical solar disc.
Spectra are shown for every third pixel, where the pixel size is 
100$''\times$100$''$.
The spectra span the entire observing bandwidth, binned into
24 frequency bands spaced 1.28~MHz apart and averaged over 0.8~MHz.
The $y$ axis ranges are (in arbitrary units):
{\em top row:} 50--4996 and 50--4620, respectively;
{\em middle row:} 100--6000 for both panels;
{\em bottom row:} 100--1800 for both panels.
Animations showing additional frames are available in the on-line journal.
The frames corresponding to the bottom two rows have fixed $y$-axis scales; those for the top row are autoscaled.
}
\label{Fig:spectra}
\end{figure}

Fig. \ref{Fig:spectra} highlights the spectroscopic capability of the 
32T by showing the spectra from representative pixels across the
entire emitting region. 
%%%%%%%%%%%%%%%%%%%%%%%%%%%%%%%%%%%%%%%%%%%%%%%%%%%%%%%%%%%%%%%%%%%%%%%%%%%%
%%%%%%%%%%%%%%%%%%%%%%%%%%%%%%%%%%%%%%%%%%%%%%%%%%%%%%%%%%%%%%%
Various interesting features can be seen here. 
Even at quiescent times, there is a significant difference in the spectral 
slope of regions close to and away from region 11057.
For both the ``narrowband'' and ``broadband'' features, the changes in 
spectral structure are confined to the vicinity of region 11057, and
to a smaller extent to the locations of other active regions visible
in the EUV image (Fig.~\ref{Fig:movie_in_time}).
The ``narrowband'' features are consistent with spectral features 
superposed on the underlying quiescent spectrum. 
In contrast, the underlying quiescent spectrum is no longer visible in 
spectra of the ``broadband'' feature, and the changes in spectral 
characteristics are significantly larger, more dynamic, and complex.

\section{Discussion and Conclusions}
\label{Sec:discussion}

At low radio frequencies, a featureless continuum due to thermal 
bremsstrahlung from a corona with optical depth approaching unity 
constitutes the bulk of the ``quiet'' Sun emission and is expected to 
vary only slowly in time \citep[e.g.,][]{Sheridan_McLean85}.
In this regime, emission that exhibits rapid spectral variations 
is dominated by plasma emission, a resonant emission mechanism
in which electrostatic Langmuir waves are converted to electromagnetic 
radiation at the local plasma frequency, $f_P$, and its harmonic, 
2$f_P$ \citep[e.g.,][]{Robinson_Cairns00}. 
This is the accepted emission mechanism for 
type~II \citep[e.g.,][]{Nelson_Melrose85} and type~III solar 
bursts \citep[e.g.,][]{Suzuki_Dulk85}.

The ability of the 32T to image and track the temporal and spectral 
evolution of coronal emission with high dynamic range and fidelity 
over a broad spectral band represents a major advance in the quality, and 
consequently the utility, of radio measurements for understanding the
dynamic solar corona.
For example, we see numerous instances of transient, non-thermal emission in 
our short observations, many of which outshine the Sun by a factor of few
and dramatically alter the $T_B$ distribution.
Though there have been earlier imaging observations of solar radio
tranisents, these investigations have been limited to a handful of
well separated, non-contiguous frequencies
due to the limited spectral sampling provided by previous 
radio interferometers.

The coherent emission mechanisms inherent at low radio frequencies 
have long been known to be a sensitive probe of electron acceleration 
in the the corona.
Given our much improved observing capability, it is not a surprise 
that in spite of significant transient behavior seen in our data, the level of 
solar activity on the day of our observations has been characterized as 
{\em low}\footnote{USAF/NOAA Solar Geophysical
`Weekly' Data} to {\em medium}\footnote{\tt{www.solarmonitor.org}}.
The SWPC ``events list'' reports nine type~III radio bursts on this day, 
none of them during the interval presented here.
The SWPC also reports seven GOES B-class and four GOES C-class 
X-flares on this day (ten and five X-ray flares, respectively in a 
24 hour period around our observations).
All of the flares for which positional information is available
(C-class) were associated with the region 11057.
The start time for one of the B-class flares coincides with the stop
time of our observations.

Based on its appearance in the frequency-time plane, the ``broadband'' 
feature present in our data (Fig. \ref{Fig:visibility_amp_phase})
resembles a weak type~III burst and shares 
similarities with the ``microbursts'' reported by \citet{Kundu86}.
A corresponding feature was seen at low signal-to-noise levels
between $\sim$80 and 
150~MHz by the Learmonth and Culgoora 
radiospectrographs,
suggesting that this feature peaked in intensity within that spectral window.
Though it appears to be a feature with a fast spectral drift rate, the 
signal-to-noise ratio and temporal sampling of the Learmonth and Culgoora
data are not sufficient for a reliable estimate of the drift rate.
Nan\c{c}ay Radioheliograph data spanning the wide 150--455~MHz frequency band show two persistent but time variable features separated by $\sim200''$ associated with region 11057 at nearby times. % (S. White, private communication).
This suggests the presence of an underlying weak type~IV emission. 
The non-thermal spectra of the region 11057 even at ``quiescent'' times, and the persistent feature seen at this location in our data are consistent with this.

Lacking an absolute flux calibration, we compute an approximate $T_B$ of
the wideband burst at the center of our observing band by assigning the 
``quiet'' Sun a $T_B$ of $\approx$7.5$\times$10$^5$$\pm$15\%~K, based on 
measurements of \citet{Lantos_Avignon75} at 169 MHz, and scaling it to the
peak of the burst emission (04:30:12.5~UT).
A measured receiver temperature of $\sim$50~K and an estimated Galactic 
background contribution of $\sim$150~K were used.
We estimate an average $T_B\sim$3.1$\times$10$^6$~K, significantly
lower than the average value of 6.3$\times$10$^7$~K reported for 
type~III bursts at 169~MHz \citep{Suzuki_Dulk85}.
The low value of $T_B$ measured highlights the ability of the 32T to make sensitive
measurements.
We note that since the angular size of the emitting region is expected to be much smaller than the 32T beam, the measured $T_B$ is expected to be considerably reduced due to beam-dilution.
Assuming a plasma emission process and the Newkirk (10$\times$Newkirk) 
model \citep{Newkirk61} for coronal electron density leads to a
coronal height of 1.7~R$_{\odot}$ (2.8~R$_{\odot}$) above the
photosphere for the source region at our central observing frequency.
A bandwidth of 30~MHz corresponds to a range of 
$\sim$0.09~R$_{\odot}$ ($\sim$0.25~R$_{\odot}$) around this coronal height.

A particularly intriguing aspect of our data is the abundance of
short-lived, ``narrowband'' emission features carpeting the frequency-time 
plane, which to our knowledge, have not been reported previously.
Though smaller in spectral and temporal extent, these features have a 
similar $T_B$, implying a coherent emission process, like plasma emission.
The existence of these emission features provides evidence for impulsive 
non-thermal energy deposition into the corona at levels lower than
previously known. 
Evaluating their contribution to the coronal heating budget 
will require building up a better understanding of the frequency of 
occurrence and energetics of such features.
Existing radio observations have been shown to be a more sensitive 
probe for such
phenomena than X-ray observations \citep[e.g.][]{Ramesh_etal2010}, and these 
observations extend the radio sensitivity advantage by another order
of magnitude.

As seen in Fig. \ref{Fig:movie_in_time}, much of the ``narrowband''
activity seen in Fig. \ref{Fig:visibility_amp_phase} maps to the
location of the region 11057.
Even the low level fluctuations in the spectrum when the Sun is in a 
relatively quiescent state (e.g., Fig. \ref{Fig:spectral_cuts}, bottom
panel; Fig.~\ref{Fig:spectra}), manifest themselves in the image plane
as flickering intensity at the location of this active region.
The occurrence of many weak X-ray flares from region 11057 around our 
observing period suggests a persistent low-level activity during our 
observations. This raises the possibility that the ``narrowband''
features are the radio signatures of the same magnetic field
rearrangement activity: numerous small reconnection events too faint to 
picked up by X-ray monitoring platforms like GOES, but with weak 
type III-like radio emission strong enough to be seen by the 32T.
This hypothesis is strengthened by the observation that the
``narrowband'' features were absent on 2010 March~29 (not shown here),
a day when no X-ray activity was reported.
There have been attempts to investigate the possibility of using 
type~III solar bursts as predictors for large solar flares in the 
past \citep{Jackson_Sheridan79, Kane81}.
The ability to characterize type~III-like activity in much greater
detail and with much higher sensitivity, demonstrated here, enhances
the prospects for investigating such a connection. 

Compared to the 32T, the MWA will provide an order of magnitude improvement 
in angular resolution, 16 times as much collecting area and 16$^2$ times 
as many baselines. 
With its unprecedented high fidelity and dynamic range polarimetric 
spectroscopic imaging capability for every spectral and temporal slice, 
and a radio quiet site providing seamless access to the RF spectrum,
the MWA will be an excellent match to the challenge of solar imaging.
Every significant advance in our capability to observe the Sun has 
revealed that the Sun and corona are more dynamic and feature-rich 
than had been thought previously.
{\it HINODE} and the {\it Solar Dynamics Observatory} ({\it SDO}) 
represent the latest in that progression.
The 32T array 
is already demonstrating that we are beginning to see features at a level
of detail not seen before, and we can justifiably expect the MWA to
similarly add to our understanding of the solar corona and the heliosphere.

\acknowledgments
This work uses data obtained from the Murchison Radioastronomy
Observatory (MRO), jointly funded by the Commonwealth Government 
of Australia and Western Australian State government.  
The MRO is managed by the CSIRO, who also provide operational support to 
the MWA. 
We acknowledge the Wajarri Yamatji people as the
traditional owners of the Observatory site. 
Support came from the U.S. National Science Foundation (grants
AST-0457585 and PHY-0835713), the Australian Research Council 
(grants LE0775621 and LE0882938), the U.S. Air Force Office of 
Scientific Research (grant FA9550-0510247), the  National Collaborative
Infrastructure Strategy, funded by the Australian federal government
via Astronomy Australia Limited, the Smithsonian 
Astrophysical Observatory, the MIT School of Science, the 
Raman Research Institute, The Australian National University, 
iVEC, the Initiative in Innovative Computing 
and NVIDIA sponsored Center for Excellence at Harvard, and the 
International Centre for Radio Astronomy Research, a Joint Venture 
of Curtin University and The University of Western 
Australia funded by the Western Australian State government.

%\bibliographystyle{apj}
%\bibliography{X13_bursts}

\end{document}